\documentclass[11pt,a4paper]{article}
\usepackage{jcappubnew}

\newcommand{\beq}{\begin{equation}}
\newcommand{\eeq}{\end{equation}}
\newcommand{\bea}{\begin{eqnarray}}
\newcommand{\eea}{\end{eqnarray}}

\newcommand{\rmi}{\mathrm{i}}
\newcommand{\rme}{\mathrm{e}}

\newcommand{\vc}[1]{{\boldsymbol #1}}
\newcommand{\mx}[1]{{\mathbf{#1}}}

\newcommand{\lh}{\left(}
\newcommand{\rh}{\right)}
\newcommand{\der}{\partial}
\renewcommand{\d}{\mathrm{d}}
\newcommand{\non}{\nonumber}
\DeclareMathSymbol{\mg}{\mathrel}{symbols}{"1D}

\newcommand{\Bnabla}{\ensuremath{\boldsymbol \nabla}}

\newcommand{\inv}{^{-1}}

\newcommand{\ga}{\alpha}
\newcommand{\gb}{\beta}

\newcommand{\gd}{\delta}
\renewcommand{\ge}{\epsilon}
\newcommand{\gve}{\varepsilon}
\newcommand{\gz}{\zeta}
\newcommand{\get}{\eta}
\newcommand{\gth}{\theta}

\newcommand{\gk}{\kappa}

\newcommand{\gn}{\nu}
\newcommand{\gx}{\xi}

\newcommand{\gs}{\sigma}

\newcommand{\gf}{\phi}

\newcommand{\gc}{\chi}

\newcommand{\gD}{\Delta}
\newcommand{\gTh}{\Theta}

\newcommand{\gP}{{\Pi}}

\newcommand{\gF}{\Phi}

\newcommand{\cO}{{\mathcal O}}
\newcommand{\cP}{{\mathcal P}}

\newcommand{\cW}{{\mathcal{W}}}

\newcommand{\f}{f_\mathrm{NL}}

\newcommand{\tW}{{\tilde W}}

\newcommand{\tgz}{{\tilde\zeta}}

\newcommand{\Bgf}{{\boldsymbol \phi}}

\newcommand{\Bgx}{{\boldsymbol \xi}}

\newcommand{\Bgth}{{\boldsymbol \theta}}

\newcommand{\Bgz}{{\boldsymbol \zeta}}

\newcommand{\Bget}{{\boldsymbol \eta}}

\newcommand{\BgP}{{\boldsymbol \gP}}

\newcommand{\bb}{{\bar b}}
\newcommand{\bc}{{\bar c}}

\newcommand{\bh}{{\bar h}}

\newcommand{\bt}{{\bar t}}

\newcommand{\bv}{{\bar v}}

\newcommand{\bA}{{\bar A}}

\newcommand{\bdm}{\begin{displaymath}}
\newcommand{\edm}{\end{displaymath}}
\newcommand{\nn}{\nonumber}
\def\be{\begin{equation}}
\def\ee{\end{equation}}

\def\vb{\bar{v}}
\def\e{\epsilon}
\def\es{\epsilon_*}
\def\hpa{\eta^{\parallel}}
\def\hpe{\eta^{\perp}}
\def\hpas{\eta^{\parallel}_*}
\def\hpes{\eta^{\perp}_*}
\def\ef{e_{1\phi}}
\def\ex{e_{1\sigma}}
\def\efs{e_{1\phi *}}
\def\exs{e_{1\sigma *}}

\title{Bispectra from two-field inflation using the long-wavelength formalism}

\author{Eleftheria Tzavara}
\author{and Bartjan van Tent}
\affiliation{Laboratoire de Physique Th\'eorique, Universit\'e Paris-Sud 11 
and CNRS, B\^atiment 210, 91405 Orsay Cedex, France}
\emailAdd{Eleftheria.Tzavara@th.u-psud.fr} 
\emailAdd{Bartjan.Van-Tent@th.u-psud.fr}

\abstract{We use the long-wavelength formalism to compute the bispectral
non-Gaussianity produced in two-field inflation. We find an exact
result that is used as the basis of numerical studies, and an
explicit analytical slow-roll expression for several classes of
potentials that gives insight into the origin and importance of the
various contributions to $\f$. We also discuss the momentum
dependence of $\f$.  Based on these results we find a simple model
that produces a relatively large non-Gaussianity. We show that the
long-wavelength formalism is a viable alternative to the standard
$\gd N$ formalism, and can be preferable to it in certain
situations.}

\begin{document}

\begin{flushright}
LPT-10-108 \\ 
\end{flushright}

\maketitle
\flushbottom

\section{Introduction}

During inflation the energy density of the Universe is assumed to be 
dominated by the potential energy of one or more scalar fields in order to 
have a sufficiently rapid expansion to solve the homogeneity problems that
plagued pre-inflationary cosmology (horizon, flatness, etc.). Very
importantly, inflation also provides the initial adiabatic density 
perturbation that generated the large-scale structure observed today.
Observations of the fluctuations in the cosmic microwave 
background radiation (CMB), in particular those made by the WMAP satellite, 
have verified the basic predictions of inflation. The most important 
observational parameters so far from the point of view of inflation have been 
the amplitude and the slope (spectral index) of the primordial power spectrum,
as well as some limits on the amount of tensor perturbations and the 
running of the spectral index.

Unfortunately this small number of observational parameters means that a 
very large number of quite different inflation models are all still
consistent with the data. To further narrow down the number of viable 
inflation models additional observables are required. A very promising
candidate is the non-Gaussianity of the primordial power spectrum. In its
most simple form this is encoded as a non-zero three-point correlator
of the CMB temperature fluctuations, or equivalently a non-zero bispectrum,
which is the Fourier (or spherical harmonic on the sphere) transform of the
three-point correlator. The quantity defined as the bispectrum divided by the
power spectrum squared is called $\f$. The current limits on this
parameter $\f$ (assuming momentum dependence of the local type, relevant for
this paper) after seven years of WMAP data are $-10<\f<74$ at 
$95\%$ confidence level \cite{Komatsu:2010fb}. The newly launched Planck
satellite \cite{:2006uk}
is expected to significantly improve these constraints, down to $1\sigma$
error bars of about $3$--$5$ (depending on the use of polarization data)
\cite{Komatsu:2001rj,Babich:2004yc}. While standard single-field slow-roll
inflation predicts 
an unobservably small value of $\f$ \cite{Maldacena:2002vr}, many other
models
predict much larger values that could be detected or ruled out by Planck.

Both supersymmetric particle theory and string theory suggest the existence
of multiple scalar fields that can influence the early Universe. If more 
than a single scalar field plays a role during inflation, 
isocurvature fluctuations will be produced in addition to the adiabatic one.
While these isocurvature fluctuations might have directly observable 
consequences in the CMB \cite{Komatsu:2010fb}, in this paper we are only 
interested in the effect of the isocurvature fluctuations on the adiabatic 
one during inflation. This effect can be important even if the isocurvature 
fluctuations disappear after inflation. The important point here is that 
while in single-field inflation the adiabatic perturbation is constant 
on super-horizon scales, this is no longer true in multiple-field inflation.
In fact, the isocurvature perturbation acts as a source for the adiabatic
perturbation on super-horizon scales and this source is multiplied by the 
$\eta^\perp$ parameter \cite{Rigopoulos:2005us}.
This $\eta^\perp$ (defined properly in the next section) is proportional
to the component of the field acceleration perpendicular to the field 
velocity. In other words, $\eta^\perp$ is non-zero if the field trajectory
makes a turn in field space. Only during such a turn will the isocurvature
mode influence the adiabatic one on super-horizon scales 
(see also \cite{Bernardeau:2002jy}).\footnote{This last 
statement is strictly only
true on a flat field manifold (trivial field metric) with standard 
kinetic terms. On a curved manifold
$\eta^\perp$ can be non-zero even for a straight field trajectory because
of the connection terms in the covariant derivatives, see 
\cite{GrootNibbelink:2001qt} and \cite{Langlois:2008mn}. In this paper a trivial
field metric will be assumed.}

There are two main ways to produce non-Gaussianity during inflation: during
and after horizon crossing of a perturbation mode. Horizon crossing
is defined as the moment when the physical wavelength $(k/a)\inv$ of the 
fluctuation becomes equal to the Hubble (or horizon) length $H\inv$, 
i.e.\ when $k=aH$ (with $k$ the wave number of the mode, $a$ the scale
factor of the Universe and $H=\dot{a}/a$ the Hubble parameter). The first
type of non-Gaussianity is produced in all inflation models, but it is 
unobservably small
(i.e.\ slow-roll suppressed), unless the model contains non-standard kinetic 
terms (higher derivatives), like DBI inflation
\cite{Alishahiha:2004eh,Langlois:2008qf} for example.
In this paper we will not consider those models, but instead focus on the
super-horizon type of non-Gaussianity. As is clear from the previous
paragraph, super-horizon non-Gaussianity can only be produced in 
multiple-field inflation models where the field trajectory makes a turn in
field space. To compute this type of non-Gaussianity
we will make use of the long-wavelength formalism developed by Rigopoulos, 
Shellard and
Van Tent \cite{Rigopoulos:2005xx, Rigopoulos:2005ae, Rigopoulos:2005us}, 
hereafter refered to as RSvT.

The main purpose of this paper is twofold. In the first place we want to
further work out, simplify and study the general analytic expression for 
$\f$ of RSvT in the case of two fields only. In the second place we want
to clarify a few remaining formal issues with the formalism, and compare
with an alternative formalism for computing $\f$, called the $\delta N$
formalism
\cite{Starobinsky:1986fxa,Sasaki:1995aw,Sasaki:1998ug,Lyth:2004gb,Lyth:2005fi}.

As will be shown, the expression for $\f$ simplifies significantly, although
a final integral remains that cannot be done analytically. As a starting point
for {\em numerical} work this expression is very useful though, and it gives 
a fully exact numerical result since no slow-roll approximation is used for the
super-horizon evolution. However, to be able to derive explicit 
{\em analytic} results, we 
have to make use of the slow-roll approximation. Using this approximation
we derive explicit analytic results for a number of classes of inflationary
potentials. Some of these have been treated before in the literature using 
the $\gd N$ formalism, but others are new.
In particular we show that models with a potential of the form
$W(\gf,\gs) = \ga \gf^p + \gb \gs^q$ with $p=q$ will never give a large $\f$ 
that persists until the end of inflation (unless inflation somehow ends
right during the turn of the field trajectory, but in that case a very
careful treatment of the transition at the end of inflation will be required).
We also present a simple model that does produce a ``large'' $\f$ of the order
of a few. The reason we choose this model is that it can be treated not only
numerically, but also analytically.

The formalism of RSvT also allows us to compute the momentum dependence of 
$\f$ due to the fact that different modes cross the horizon at different
times. This effect has usually been ignored in the literature
where it was often assumed that $\f$ is momentum-independent (see e.g.
\cite{Vernizzi:2006ve}), although recently people have started looking
into this \cite{Byrnes:2009pe,Byrnes:2010ft}. Here we compute this 
momentum dependence in an exact way and show that, depending on the model, 
it can lead to relative effects of order 10\% even within the range 
of momenta that are observable by Planck.

We have extended the formalism of RSvT with an exact treatment of the
second-order source term at horizon crossing. While negligibly small in
the models we consider, the inclusion of this term allows for an exact
analytic comparison with the $\gd N$ formalism. We find that
our analytic slow-roll results agree exactly with those derived
using the $\delta N$ formalism, where available. 
For models where slow roll breaks down long after horizon crossing and which 
have to be treated numerically we also find excellent agreement. 

Apart from providing an 
alternative way of computing the bispectrum, which is always useful, the
long-wavelength formalism provides a number of advantages compared to the 
$\gd N$ formalism.
Very importantly, the long-wavelength formalism allows for a 
simple physical interpretation of the different terms, showing the contributions
from adiabatic and isocurvature modes and making clear why some of them can 
become big and others cannot. While we do not pursue this in the present paper,
the formalism also provides the solution for the second-order isocurvature
perturbation and hence the isocurvature bispectrum could be computed as
easily as the adiabatic one.

Many people have worked on non-Gaussianity, both predictions from
inflation and estimators for CMB observations. The reduced bispectrum
has been given in \cite{Creminelli:2005hu} for the equilateral type,
in \cite{Komatsu:2001rj,Babich:2004yc} for the local type and in
\cite{Senatore:2009gt} for the orthogonal type of the primordial
bispectrum. The different shapes were studied in detail in 
\cite{Babich:2004gb,Fergusson:2008ra}.
Bispectrum estimators were developed in 
\cite{Komatsu:2003iq,Creminelli:2005hu,Yadav:2007rk,Fergusson:2009nv,
Bucher:2009nm}.
All kinds of inflationary models have been studied as
well. For instance, we have learned that single-field inflation models
cannot produce large non-Gaussianity \cite{Maldacena:2002vr}, unless
some non-trivial potential is used \cite{Chen:2006xjb} or higher
derivative contributions are introduced as for the Dirac-Born-Infeld
action
\cite{Alishahiha:2004eh,Silverstein:2003hf,Mizuno:2009cv,Mizuno:2010ag}
or K-inflation \cite{Chen:2006nt,Chen:2009bc}. The study of the
effective theory of inflation has also turned out to be very fruitful
\cite{Cheung:2007st}. These models were extended to incorporate
multiple fields in
\cite{Langlois:2008qf,Arroja:2008yy,Mizuno:2009cv,Cai:2009hw,Senatore:2010wk}. 
Large non-Gaussianity can also be produced at the end of inflation
\cite{Lyth:2005qk,Bernardeau:2004zz,Barnaby:2006km,Enqvist:2004ey,
Enqvist:2005qu,Jokinen:2005by}
or after inflation, in models with varying inflaton decay rate
\cite{Zaldarriaga:2003my} and in curvaton models
\cite{Bartolo:2003jx,Enqvist:2005pg,Ichikawa:2008iq,Malik:2006pm,Sasaki:2006kq,
Huang:2008zj}.

Large scale evolution of perturbations during inflation up to second
order became possible through their consistent gauge invariant
definition \cite{Malik:2003mv,Rigopoulos:2003ak,Langlois:2005qp}. 
Within the $\delta N$
formalism several authors have investigated the bispectra of specific
multiple field inflation models
\cite{Seery:2005gb,Kim:2006te,Battefeld:2006sz,Battefeld:2007en,
Langlois:2008vk,Cogollo:2008bi}. Two-field models, being easier to deal with, 
have gained popularity
though. Vernizzi and Wands studied the double field sum potential
\cite{Vernizzi:2006ve}, while the double product potential was studied
in \cite{Choi:2007su}. Conditions for large non-Gaussianity were found in
\cite{Byrnes:2008wi}.

The paper is organized as follows. In section \ref{basic} we summarize
the long-wavelength formalism of RSvT and define the various quantities 
used in the paper. Here we also describe the second-order source term,
which is a new extension of the formalism. At the end of the section
we give a brief overview of the $\gd N$ formalism for comparison
purposes. In section \ref{general} we work out the general expression
for $\f$ in the case of two fields. We also compute the momentum dependence
that arises in the case that not all scales cross the horizon at the same
time. These expressions, derived without using any super-horizon slow-roll 
approximation, are one of the main results of the paper and the starting point 
for our numerical analyses. In order to find completely explicit analytic
expressions for $\f$, however, we do need to assume slow roll, as well as some 
conditions on the potential. This is treated in section \ref{secSlowRoll},
where we also compare our analytic results with those obtained using the
$\gd N$ formalism, for as far as the latter exist. 
In section \ref{numerical} we use the two-field quadratic potential in
order to compare our exact numerical results with those of the 
$\delta N$-formalism. We also present a simple potential that can produce 
an $\f$ of the order of a few, which falls into the category of potentials
that can be treated analytically, thus allowing us to test our results.
We conclude in section~\ref{secConclusion}. Finally, in the appendices we 
give supplementary information on the basis in field space that we use, 
compute the second-order source term, comment on some gauge and formal 
issues, and provide several intermediate steps of our calculations. Note that 
\ref{appBasis} introduces in particular a small improvement of the basis 
defined in \cite{GrootNibbelink:2001qt} that makes it more convenient for 
numerical calculations during periods when the fields oscillate.

\section{Basic equations and definitions}
\label{basic}

This section sets up the starting point for the work in the following sections.
It is mostly a summary of the long-wavelength formalism and its results as 
presented in
\cite{Rigopoulos:2005us}, although the part about the second-order source
term in section~\ref{secNonlinear} and some results on the Green's functions in
section~\ref{secGreen} are new. Section~\ref{secNonlinear} describes the 
non-linear equations for the perturbations, section~\ref{secGreen} shows how 
to solve
them using Green's functions, and section~\ref{secCorrelators} gives the
formal expressions for the two and three point correlation functions of the
perturbations, or rather their Fourier transforms, the power spectrum and the 
bispectrum. It is the latter that will be worked out in great detail in the
rest of the paper.
Finally in section~\ref{dnform} we give a brief overview of the $\gd N$ 
formalism for comparison purposes later in the paper.

\subsection{Non-linear equations}
\label{secNonlinear}

In the long-wavelength formalism space-time is described by the
long-wavelength metric \cite{Salopek:1990re}
\be
\d s^2=-N^2(t,\vc{x})\d t^2+a^2(t,\vc{x})\d\vc{x}^2,
\label{metric}
\ee
where the lapse function $N(t,\vc{x})$ defines the time
slicing and $a(t,\vc{x})$ is the space dependent scale factor. The 
Hubble parameter is defined as $H\equiv \partial_t\ln{a}/N$.
In order to simplify superhorizon calculations we choose to work in a flat
gauge $NH=1$, that is we choose time slices in which the expansion of the 
universe is homogeneous and the time variable coincides with the number of
e-folds $t=\ln{a}$. 

On the matter side we assume in this paper two scalar fields with a trivial
field metric, although the formalism can in principle deal with an arbitrary 
number of scalar fields living on an arbitrary field manifold. The 
energy-momentum tensor for the two fields $\gf^A$ ($A,B = 1,2$) is
\be
T_{\mu\nu}=\gd_{AB}\partial_{\mu}\phi^A\partial_{\nu}\phi^B-g_{\mu\nu}
\left(\frac{1}{2}\gd_{AB}\partial^{\lambda}\phi^A\partial_{\lambda}\phi^B
+W\right),
\ee
where $W$ is the potential. The Einstein summation convention is assumed
throughout this paper. We also define the derivative of the
fields with respect to proper time as $\Pi^A\equiv\partial_t\phi^A/N$,  
with length $\Pi$. 

The long-wavelength formalism corresponds to the leading-order
approximation of the spatial gradient expansion 
(see \cite{Salopek:1990jq,Tanaka:2006zp} and references therein). 
In this expansion all quantities are expanded in terms of a small parameter 
$1/(HL)$, where $L$ is the characteristic physical length scale of the 
perturbations (i.e. proportional to $a$). The leading-order approximation 
of the spatial gradient expansion is equivalent to neglecting the $k^2$ term 
(which comes from the second-order spatial gradient and is of order 
$\cO(1/(HL)^2)$) with respect to the $\cO(1)$ terms in the 
equation for the perturbation modes. Because of the very rapid growth of $a$ 
during inflation, this is in principle a well-justified approximation from 
just a few e-folds after horizon crossing of the perturbation mode under 
consideration, when the decaying mode will have disappeared.
However, as pointed out in \cite{Leach:2001zf,Takamizu:2010xy},
if slow roll is broken at horizon crossing and for some e-folds afterwards,
a cancellation of the $\cO(1)$ terms can cause the decaying mode to remain 
important during this period. In those papers it was shown that 
for single-field inflation there may be an enhancement of
the curvature perturbation both at first and second order due to the effect 
of the $k^2$ term even on super-horizon scales, if the decaying mode has 
not yet vanished.
In this paper we will assume slow roll to hold around horizon crossing,
so that the decaying mode will quickly disappear, and the long-wavelength 
approximation is valid on super-horizon scales.

The Einstein and field equations are given by \cite{Rigopoulos:2005xx}
\bea
&&H^2 = \frac{\kappa^2}{3}\left(\frac{\Pi^2}{2}+W\right),
\qquad\qquad
\dot{H}  =-\frac{\kappa^2 \Pi^2}{2H}, \non\\
&&\dot{\Pi}^A  = -3\Pi^A-\frac{W^{,A}}{H},
\qquad\qquad\ \ 
\der_i H  = - \frac{\gk^2}{2} \Pi_A \der_i \gf^A,
\label{fieldeq}
\eea
with $\kappa^2\equiv8\pi G=8\pi/m_{pl}^2$ and $W_{,A}\equiv\der W/\der\gf^A$.
While the first three equations look like background equations, they are 
actually fully non-linear since $H(t,\vc{x})$, $\gf^A(t,\vc{x})$, and
$\Pi^A(t,\vc{x})$ defined above are functions of both time and space.
We define an orthonormal
basis ${e_m^A}$ in field space (see \ref{appBasis} for details) 
through the field velocity, $e_1^A\equiv\Pi^A/\Pi$, and successively 
higher-order time derivatives of the fields \cite{GrootNibbelink:2001qt}. 
This basis allows us to 
easily distinguish effectively single-field effects with $m=1$
from truly multiple-field effects with $m \geq 2$. 
The local slow-roll parameters then take the form 
\bea
\ge(t,\vc{x})  &\equiv& - \frac{\dot{H}}{H},
\qquad\qquad\qquad 
\hpa(t,\vc{x}) \equiv \frac{e_{1A}\dot{\Pi}^A}{\Pi},
\qquad\qquad\;
\hpe(t,\vc{x}) \equiv \frac{e_{2A}\dot{\Pi}^A}{\Pi}, \non\\
\chi(t,\vc{x}) &\equiv& \tilde{W}_{22}+\e+\hpa,\qquad\ 
\gx^\parallel(t,\vc{x}) \equiv \frac{e_{1A}\ddot{\Pi}^A}{\Pi} - \ge\hpa,
\qquad\!\!
\gx^\perp(t,\vc{x}) \equiv \frac{e_{2A}\ddot{\Pi}^A}{\Pi} - \ge\hpe,
\label{srvar}
\eea
where $\tilde{W}_{mn}\equiv W_{mn}/(3H^2)$ 
and $W_{mn}\equiv e_m^Ae_n^BW_{,AB}$. 
Throughout this paper the indices $l, m, n$ will indicate components in the
basis defined above, taking the values 1 and 2.
The correction term in the expressions
for $\gx^\parallel$ and $\gx^\perp$ comes about because the proper definition
is $\gx^A \equiv ([(1/N) \der_t]^2 \gP^A)/(H^2\gP)$ and $NH=1$. We note here 
that we have not made any slow-roll approximations so far; the above 
quantities should be viewed as short-hand notation and can be large.
We also give the time derivatives of the slow-roll parameters,
	\bea\label{SRder}
	\dot{\ge} & = & 2 \ge ( \ge + \get^\parallel ),\qquad
	\dot{\get}^\parallel = \gx^\parallel + (\get^\perp)^2 
	+ (\ge - \get^\parallel) \get^\parallel, \qquad
	\dot{\get}^\perp = \gx^\perp + (\ge - 2\get^\parallel) \get^\perp,
	\non\\
	\dot{\gc} & = & \ge \get^\parallel + 2 \ge \gc - (\get^\parallel)^2
	+ 3 (\get^\perp)^2 + \gx^\parallel + \frac{2}{3} \get^\perp \gx^\perp
	+ \tilde{W}_{221},
	\eea
and of the unit vectors,
\be
\dot{e}_1^A=\hpe e_2^A,\qquad\qquad
\dot{e}_2^A=-\hpe e_1^A.
\label{dere}
\ee
Here we have defined $\tW_{lmn} \equiv (\sqrt{2\ge}/\gk)W_{lmn}/(3H^2)$.
Hence the tilde has a different meaning in the case of two and of three 
derivatives, but it is these specific combinations that always appear in
the equations. 

Our main variable describing the perturbations is \cite{Rigopoulos:2005xx}
\be
\zeta_i^m\equiv\delta_{m1}\partial_i\ln{a}-\frac{\kappa}{\sqrt{2\epsilon}}
\left(e_{mA}\partial_i\phi^A\right),
\label{zdef}
\ee
with $m=1$ the adiabatic component and $m=2$ the isocurvature one.
The non-linear quantity $\gz_i^m$ has been constructed to transform as a scalar 
under changes of time slicing on long wavelengths. In the single-field case
and when linearized it is just the spatial gradient of the well-known
curvature perturbation $\gz$.
Of course in the gauge we have chosen $\der_i \ln a = 0$ and the first term
in the expression for $\gz_i^m$ disappears.
The exact evolution equations for $\zeta_i^m$ and its time
derivative $\theta_i^m\equiv\partial_t\left(\zeta_i^m\right)$ are
\cite{Rigopoulos:2005us}
\bea
\dot{v}_{ia}(t,\vc{x})+A_{ab}(t,\vc{x})v_{ib}(t,\vc{x})=0,
\qquad\mathrm{where}\quad
v_{ia}\equiv\left(\zeta_i^1,\zeta_i^2,\theta_i^2\right),
\label{nonlineareqlw}
\eea
so $a,b = 1,2,3$. We have simplified the system by omitting the time 
derivative of the adiabatic component, $\gth_i^1$, since it is given by 
$\theta_i^1=2\hpe\zeta_i^2$, valid fully non-linearly 
\cite{Rigopoulos:2005us}. The matrix $A_{ab}$ in the two-field case is
\be
\label{Amat}
\mx{A} = 
\left(\begin{array}{ccc}
0 & -2 \get^\perp & 0\\ 
0 & 0 & -1\\
0 & 3\gc+2\ge^2+4\ge\get^\parallel+4(\get^\perp)^2+\gx^\parallel 
& 3+\ge+2\get^\parallel
\end{array}\right).
\ee

To solve this equation, we expand the system as an infinite hierarchy of linear
inhomogeneous perturbation equations. To first and second order we obtain
\bea
\dot{v}_{ia}^{(1)}+A_{ab}^{(0)}(t)v_{ib}^{(1)}&=&b_{ia}^{(1)}(t,\vc{x}),
\label{evol1}\\
\dot{v}_{ia}^{(2)}+A_{ab}^{(0)}(t)v_{ib}^{(2)}&=&
-A_{ab}^{(1)}(t,\vc{x}) v_{ib}^{(1)} + b_{ia}^{(2)}(t, \vc{x} ).
\label{evol2}
\eea
Here the source terms $b^{(1)}$ and $b^{(2)}$ have been added to describe
the influence of the short-wavelength modes on the long-wavelength system
given in (\ref{nonlineareqlw}), providing the necessary initial conditions.
$A_{ab}^{(1)}$ is found by perturbing the exact $\mx{A}$ matrix, giving
$A_{ab}^{(1)}(t,\vc{x})=\bar{A}^{(0)}_{abc}(t)v_c^{(1)}(t,\vc{x})$. 
The explicit form of $\mx{\bar{A}}$ is given in (\ref{Abar}), where we
have dropped the superscript $(0)$ for notational convenience.
We have also defined
$v_c^{(1)}\equiv \partial^{-2}\partial^iv_{ic}^{(1)}$. 

The source term $b_{ia}^{(1)}$ can be expressed in terms of the linear mode 
function solutions $X_{am}^{(1)}$, using a window function
$\mathcal{W}(k)$ which guarantees that short wavelengths are cut out to get
only the contribution to the long-wavelength system,
\be
b_{ia}^{(1)}=\int\frac{\d^3\vc{k}}{(2\pi)^{3/2}}\dot{\mathcal{W}}(k)X_{am}^{
(1)}(k) \hat { a }
_m^{\dagger}(\vc{k})\rmi k_i \rme^{\rmi\vc{k}\cdot\vc{x}}+\mathrm{c.c.}.
\ee
The quantum creation ($\hat{a}^{\dagger}_m$) and conjugate annihilation 
($\hat{a}_m$) operators satisfy the usual commutation relations. The linear
mode solutions can be determined exactly numerically, or analytically within
the slow-roll approximation (which, as observations indicate, seems to be a
very good approximation at horizon crossing). See section~\ref{secGreen} for
explicit expressions for these $X_{am}^{(1)}$ as well as for the window 
function $\mathcal{W}$.

The source term $b_{ia}^{(2)}$ was either neglected by RSvT 
\cite{Rigopoulos:2005us} because it is small, or in earlier papers 
(e.g. \cite{Rigopoulos:2005ae}) approximated by perturbing
$X_{am}^{(1)}$, which turned out not to be a good approximation.
While this contribution to $\f$ is indeed small, here we compute it
explicitly in order to allow for an exact comparison with known results in the
literature. We find that it can be expressed by means of the window 
function as 
\bea\label{secondsource}
 b_{ia}^{(2)}= && \int\frac{\d^3\vc{k}}{(2\pi)^{3/2}}\int\frac{\d^3\vc{k'}}{(2\pi)^{3/2}}
\dot{\mathcal{W}}(\mathrm{max}(k',k))\nn\\
&&\times \Bigg\{L_{abc}(t)X_{bm}^{(1)}(k',t)X_{cn}^{(1)}(k,t)
\hat { a }_m^{\dagger}(\vc{k'})
\hat { a }_n^{\dagger}(\vc{k})\rmi (k'_i+k_i)
\rme^{\rmi(\vc{k'}+\vc{k})\cdot\vc{x}}\nn\\
&&\ \ \ \ \  +N_{abc}(t)X_{bm}^{(1)}(k',t)X_{cn}^{(1)}(k,t)
\hat { a }_m^{\dagger}(\vc{k'})
\hat { a }_n^{\dagger}(\vc{k})\rmi k_i\rme^{\rmi(\vc{k'}+\vc{k})\cdot\vc{x}}
+\mathrm{c.c.}\Bigg\},
\eea
where the derivative of the window function peaks at the scale that exits 
the horizon last. 
We have split $b_{ia}^{(2)}$ into a local part proportional to $L_{abc}$ and 
a non-local part proportional 
to $N_{abc}$. In order to find these factors,
we generalize the results of Maldacena \cite{Maldacena:2002vr} to
multiple fields.  This is similar to the work done in
\cite{Seery:2005gb}, but due to the different definitions and gauge
choices used in that paper, we found it easier to rederive the results
from scratch. Maldacena computed the third-order action for $\zeta$ 
in the uniform energy density gauge. 
In order to calculate the three-point correlation function he performed a 
redefinition of $\zeta$ to remove terms in the action which are proportional
to the equations of motion.
Our generalization of this calculation can be found in \ref{appSecondSource},
while the explicit expressions for the components of $L_{abc}$ and $N_{abc}$
are given in section~\ref{secGreen}.

\subsection{Green's functions}
\label{secGreen}

Equations (\ref{evol1}) and (\ref{evol2}), together with the initial condition
$v_{ia}\left(t\rightarrow-\infty\right)=0$, can be solved using a simple Green's
function $G_{ab}(t,t')$. In matrix notation it satisfies
\cite{Rigopoulos:2005us,Rigopoulos:2005ae}
	\beq\label{Greeneqmot}
	\frac{\d}{\d t} \mx{G}(t,t') + \mx{A}(t) \mx{G}(t,t') = \mx{0},
	\qquad\qquad
	\mx{G}(t,t) = \mx{1}.
	\eeq
Starting from this equation, to lighten the notation, when we write $\mx{A}$ 
we actually mean $\mx{A}^{(0)}$, i.e.\ the matrix in (\ref{Amat}) with all 
local slow-roll parameters replaced by their background version that depends 
on time only.
Looking at this equation of motion and its initial condition, we see that the
solution can be written as
	\beq\label{GisFFinv}
	\mx{G}(t,t') = \mx{F}(t) \mx{F}\inv(t'),
	\eeq
where $\mx{F}(t)$ satisfies the same equation of motion (\ref{Greeneqmot}) as
$\mx{G}(t,t')$ with an arbitrary initial condition. From this we immediately
derive that
	\beq
	\frac{\d}{\d t'} \mx{G}(t,t') - \mx{G}(t,t') \mx{A}(t') = \mx{0}.
\label{grstar}
	\eeq

The solution of (\ref{evol1}) and (\ref{evol2}) can now be written as the 
time integral of $G_{ab}$ contracted with the terms on
the right-hand side of these equations:
\be\label{via1sol}
v_{ia}^{(1)}(t,\vc{x})=\int\frac{\d^3\vc{k}}{(2\pi)^{3/2}}v_{am}(k,
t)\hat { a }
_m^{\dagger}(\vc{k})\rmi k_i \rme^{\rmi\vc{k}\cdot\vc{x}}+\mathrm{c.c.},
\ee
with
\be
v_{am}(k,t)=\int_{-\infty}^t\d
t'G_{ab}(t,t')\dot{\mathcal{W}}(k,t')X_{bm}^{(1)}(k,t')
\ee
and
\be
\label{u2hor}
 v_{ia}^{(2)}(t,\vc{x})=-\int_{-\infty}^t\!\!\!\d
t'G_{ab}(t,t')\bar{A}_{bcd}(t')v_{ic}^{(1)}(t',\vc{x})v_{d}^{(1)}(t',\vc{x})+
\int_{-\infty}^t\!\!\!\d t'G_{ab}(t,t')b_{ib}^{(2)}(t',\vc{x}).
\ee
As before, $v_d^{(1)}\equiv \partial^{-2}\partial^iv_{id}^{(1)}$.

Written in components (\ref{Greeneqmot}) gives the following equations
for the Green's functions:
	\bea\label{Greeneqmot2f}
	\frac{\d}{\d t} G_{1x}(t,t') & = & 2 \get^\perp(t) G_{2x}(t,t'),
	\non\\
	\frac{\d}{\d t} G_{2x}(t,t') & = & G_{3x}(t,t'),
	\\ 
	\frac{\d}{\d t} G_{3x}(t,t') & = & - A_{32}(t) G_{2x}(t,t')
	- A_{33}(t) G_{3x}(t,t'),
	\non\\
	G_{ab}(t,t) & = & \gd_{ab}.
	\non
	\eea
We can also rewrite this as a second-order differential equation for $G_{2x}$:
	\beq\label{G22eq}
	\frac{\d^2}{\d t^2} G_{2x}(t,t') + A_{33}(t) \frac{\d}{\d t}
	G_{2x}(t,t') + A_{32}(t) G_{2x}(t,t') = 0.
	\eeq
For the derivatives with respect to $t'$ we find:
	\bea\label{Greeneqmot2ftp}
	\frac{\d}{\d t'} G_{x2}(t,t') & = & -2 \get^\perp(t') \gd_{x1}
	+ A_{32}(t') G_{x3}(t,t'),
	\non\\ 
	\frac{\d}{\d t'} G_{x3}(t,t') & = & - G_{x2}(t,t') 
	+ A_{33}(t') G_{x3}(t,t').
	\eea

The solutions for the $x=1$ components of (\ref{Greeneqmot2f}) are simple: 
$G_{11} = 1$, $G_{21} = G_{31} = 0$. 
To find the solutions for the $x=2,3$ components 
we assume that we have found a solution $g(t)$ that satisfies (\ref{G22eq}). 
Then a second, independent, solution is given by
	\beq\label{deffY}
	f(t) = g(t) \int^t \d\bt \: Y(\bt),
\qquad\qquad 
	Y(t) \equiv \frac{1}{g^2(t)}e^{- \int^t \d\bt A_{33}(\bt)}
	= \frac{1}{g^2(t)} \frac{\mathrm{e}^{-3t}}{H(t) \ge(t)}.
	\eeq
Hence
	\bea
	 &&G_{23}(t,t') =\frac{1}{g(t') Y(t')} f(t)
	- \frac{f(t')}{g^2(t') Y(t')}g(t),
	\\
	 &&G_{22}(t,t')= \lh \frac{\dot{g}(t') f(t')}
	{g^3(t') Y(t')} + \frac{1}{g(t')} \rh g(t)- \frac{\dot{g}(t')}{g^2(t') Y(t')}f(t)
=\frac{g(t)}{g(t')}-\frac{\dot{g}(t')}{g(t')}G_{23}(t,t'),\qquad
	\eea
and
	\be
	 G_{33}(t,t')= \frac{\dot{g}(t)}{g(t)} \, G_{23}(t,t')
	+ \frac{g(t) Y(t)}{g(t') Y(t')},\quad
	G_{32}(t,t') = \frac{\dot{g}(t)}{g(t)} \, G_{22}(t,t')
	- \frac{\dot{g}(t')}{g(t')} \frac{g(t) Y(t)}{g(t') Y(t')}.
	\ee	
Of course $G_{13}(t,t') = 2 \int_{t'}^t \d\bt \, \get^\perp(\bt)
G_{23}(\bt,t')$ and $G_{12}(t,t') = 2 \int_{t'}^t \d\bt \, \get^\perp(\bt)
G_{22}(\bt,t')$.
For exact calculations the Green's functions will be determined numerically,
but in an approximate slow-roll treatment we can sometimes find analytic
solutions, see section~\ref{secSlowRoll}.

For the linear mode solutions at horizon crossing, $X_{am}^{(1)}$, we will
assume in this paper the analytic slow-roll solutions determined in
\cite{GrootNibbelink:2001qt}. Observations of the spectral index indicate
that slow roll is a good approximation {\em at} horizon crossing. Note 
however that, with the exception of section~\ref{secSlowRoll}, we do not 
assume slow roll to hold {\em after} horizon crossing. Moreover, the assumption
of slow roll at horizon crossing is not a requirement to compute these
linear solutions, we could just as well numerically compute the linear mode 
solutions exactly.
For the window function used in the
calculation of the linear solution we take a step function, see
\cite{Rigopoulos:2005xx,Rigopoulos:2005us}, so that its time
derivative is a delta function: $\dot{\cW} = \gd(k c/(a H \sqrt{2}) - 1)$, where
$c$ is a constant of the order of a few, e.g.\ $c=3$. Then
\beq
v_{am}(t) = G_{ab}(t,t_*) X_{bm}^{(1)}(t_*) \gTh(t-t_*),
\eeq
where the step function $\gTh(x)$ equals 1 for $x\geq 0$ and 0 for $x<0$. 
The time $t_*$ is defined by $aH=kc/\sqrt{2}$, i.e.\ a time slightly after
horizon crossing when we have entered the long-wavelength regime. 
While results right at $t_*$ of course depend on the details of the window
function, a few e-folds later any dependence on $\cW$ has disappeared.
Moreover, under the assumption of slow roll at horizon crossing, all
quantities change very little between horizon crossing and $t_*$, so that
final results do not depend on the choice of $c$ and $t_*$ can be taken
equal to the horizon-crossing time determined from $k=aH$ in the final
expressions.
Defining $\gamma_*$ as $\gamma_* \equiv 
- \gk H_*/(2 k^{3/2} \sqrt{\ge_*})$, where the subscript $*$ means evaluation 
at $t=t_*$, the matrix $\mx{X}^{(1)}(t_*)$ is given by 
$X_{11}^{(1)}(t_*) = X_{22}^{(1)}(t_*) = \gamma_*$,
$X_{32}^{(1)}(t_*) = -\gc_* \gamma_*$, the other components being zero
\cite{Rigopoulos:2005us}. 
Hence we have
\bea
&v_{11} =  \gamma_* \gTh(t-t_*), &	v_{12}(t) =  \gamma_* \lh G_{12}(t,t_*) 
- \gc_* G_{13}(t,t_*) \rh \gTh(t-t_*),\nn\\
	&v_{21} =  0, &  v_{22}(t) =  \gamma_* \lh G_{22}(t,t_*) 
- \gc_* G_{23}(t,t_*) \rh \gTh(t-t_*),\nn\\
	&v_{31} =  0, & 	v_{32}(t) =  \gamma_* \lh G_{32}(t,t_*) 
- \gc_* G_{33}(t,t_*) \rh 	\gTh(t-t_*).\label{vamGreenrel}
\eea
We also define the short-hand notation $\bv_{am}$ by 
$v_{am}(t) = \gamma_* \gTh(t-t_*)\bv_{am}(t)$.

For the second-order horizon-crossing solutions we find from (\ref{z1fg}) and
(\ref{z2f}) (see \ref{appSecondSource}) that the slow-roll matrices $L_{abc}$ 
and $N_{abc}$ have elements satisfying
\bea
&&L_{111*} =\e_*+\hpa_*,\qquad\qquad\qquad\qquad L_{122*}=-\lh\e_*+\hpa_*-\chi_*\rh,\nn\\
&&L_{211*} =\hpe_*,\qquad\qquad\qquad\qquad\ \ \ \ \ \ L_{222*}=\hpe_*,\nn\\
&&L_{112*}+L_{121}=2\hpe_*,
\ \ \ \ \qquad\qquad\ \ \  N_{112*}+N_{121*}=-2\hpe_*,\nn\\
&&L_{212*}+L_{221*}=2\lh\e_*+\hpa_*-\chi_*\rh,
\  N_{212*}+N_{221*}=\chi_*,
\eea
with the other elements of $N_{abc}$ being zero. As explained in the appendix,
a slow-roll approximation which expresses $\gth^2$ in terms of $\gz^2$ has 
been used. This means in particular that the subscripts $a,b,c$ only take the 
values 1 and 2, but not 3. However, for consistency in the notation, we
will define here all entries of $L_{abc}$ and $N_{abc}$ to be equal to zero 
if one or more of the indices are equal to 3.

\subsection{Two and three point statistics}
\label{secCorrelators}

So far we have used time slices on which the expansion of the universe is
homogeneous ($\der_i \ln a = 0$), since it simplifies super-horizon 
calculations. However, to make contact with the proper gauge-invariant
expression for $\gz$ it turns out to be 
necessary to change to uniform energy density time slices 
($\partial_i\rho=0$). On such slices 
the adiabatic perturbation variable has the simple form
\be
\tilde{\zeta}_i^1=\partial_i\ln{a}.
\label{redefinedzetagauge}
\ee
At first and second order the relation between the adiabatic component
of $\gz_i$ in the two gauges is (see \cite{Rigopoulos:2005us} and the
discussion in \ref{appGauge})
\bea
\label{gaugetrans}
\tilde{\zeta}_i^{(1)1}(t)&=&\zeta_i^{(1)1}(t),\\
\tilde{\zeta}_i^{(2)1}(t)&=&\zeta_i^{(2)1}(t)+2\hpe\zeta^{(1)1}\zeta_i^{(1)2},
\eea
where again $\zeta^{(1)1}\equiv \partial^{-2}\partial^i\zeta_i^{(1)1}$.
Indeed one can show that not only do we end up with a total gradient through 
this gauge transformation, we also obtain the gauge-invariant quantity 
corresponding to the curvature perturbation $\zeta$:
in the flat gauge and for superhorizon scales one can show that \cite{Tzavara:2011hn} 
(see also \cite{Langlois:2005qp} for 
the energy density definition of $\zeta_i$)
\be
\zeta_i^{(2)1}=\partial_i\zeta^{(2)1}-\zeta^{(1)1}\dot{\zeta}_i^{(1)1}.
\ee 
The second term on the right-hand side cancels exactly the 
gauge transformation term and we are left with the space gradient of the 
gauge-invariant quantity.

From (\ref{via1sol}) we see that the Fourier coefficients of 
$\gz^{(1)1}(\vc{x},t)$ 
are given by $\gz^{(1)1}_{\vc{k}}(t) = v_{1m}(k,t) ( \hat{a}_m^{\dagger}(\vc{k})
+ \hat{a}_m(-\vc{k}) )$. Hence the power spectrum, that is the two-point
correlator of the Fourier coefficients, is
\be
\langle\tilde{\zeta}^{(1)1}_{\vc{k_1}} \tilde{\zeta}^{(1)1}_{\vc{k_2}}\rangle
 = \delta^3(\vc{k_1}+\vc{k_2}) v_{1m}(k_1, t)v_{1m}(k_1, t).
\ee
Conventionally (see e.g. \cite{Lyth:1998xn,Bartolo:2004if}) a quantity 
$\cP_\gz$, also called the 
power spectrum of $\gz$, is defined to remove the overall delta function and 
the factor $1/k^3$ coming from the $v_{1m}$ (see (\ref{vamGreenrel})), as
follows:
\beq
\mathcal{P}_{\zeta}(k,t) \equiv \frac{k^{3}}{2\pi^2} \, 
v_{1m}(k, t) v_{1m}(k, t).
\label{power}
\eeq
The scalar spectral index is then defined as
\be
n_{\zeta} - 1 \equiv\frac{\d\ln{\mathcal{P}_{\zeta}}}{\d\ln{k}}
= \frac{\d\ln{\mathcal{P}_{\zeta}}}{\d t_*}\frac{\d t_*}{\d\ln{k}}
=\frac{\d\ln{\mathcal{P}_{\zeta}}}{\d t_*}\frac{1}{1-\e_*},
\label{spectral}
\ee
where we used that $k=aH\sqrt{2}/c$ and $\dot{H}=-\e H$ for our gauge where
$t=\ln a$.

By combining the different permutations of
$\langle\tilde{\zeta}_{\vc{k_1}}^{(2)1}\tilde{\zeta}_{\vc{k_2}}^{(1)1}
\tilde{\zeta}_{\vc{k_3}}^{(1)1}\rangle$ of the Fourier components of the 
linear and second-order 
adiabatic solutions (first subtracting the average of $\tgz^{(2)1}(\vc{x},t)$ 
to get rid of the divergent part), we find the 
bispectrum\footnote{In the literature (e.g. \cite{Maldacena:2002vr}) one 
often sees a factor $(2\pi)^3$ in front of the bispectrum (as well as in 
front of the power spectrum). This is due to a different definition of the 
Fourier transform. We use the convention where both the Fourier transform
and its inverse have a factor $(2\pi)^{-3/2}$.}
\cite{Rigopoulos:2005us}
\bea
\langle \tilde{\zeta}_{\vc{k_1}}^{1} \tilde{\zeta}_{\vc{k_2}}^{1}
\tilde{\zeta}_{\vc{k_3}}^{1}\rangle^{(2)} & = & 
(2\pi)^{-3/2}\delta^3(\sum_s\vc{k_s})\left[f(k_1,k_2)+f(k_1,k_3)+f(k_2,k_3)
\right]\non\\
& \equiv & (2\pi)^{-3/2}\delta^3 (\sum_s\vc{k_s})B_{\zeta}(k_1,k_2,k_3),
\eea
where
\bea
 f(k,k') \equiv v_{1m}(k) v_{1n}(k') \Bigg( && \get^{\perp} 
v_{2m}(k) v_{1n}(k')+
\frac{1}{2}G_{1a}(t,t_{k'})M_{abc}(t_{k'})X_{bm}(k,t_{k'})
X_{cn}(k',t_{k'})\non\\
&& - \frac{1}{2}\int_{-\infty}^t\d t'\,G_{1a}(t,t')\bA_{abc}v_{bm}(k)v_{cn}(k')\Bigg)
+ {k} \leftrightarrow {k}',
\label{fkk}
\eea
where $k'$ refers to the scale that exits the horizon last and 
$M_{abc} \equiv L_{abc}+N_{abc}$.
Finally we introduce the parameter $\f$, basically defined as the bispectrum
divided by the power spectrum squared, which gives a relative measure of the
importance of non-Gaussianities of the bispectral type (see 
e.g.\ \cite{Maldacena:2002vr, Vernizzi:2006ve}):
\bea\label{fNL_start}
-\frac{6}{5}\f & \equiv & \frac{B_{\zeta
} (k_1,k_2,k_3)} {\frac{2\pi^2}{k_1^3}\mathcal{P}_{\zeta}(k_1)\frac{2\pi^2}{k_2^3}\mathcal{P}_{\zeta}(k_2)
+(k_2\leftrightarrow k_3) + (k_1\leftrightarrow k_3)} \non\\
& = & \frac{f({k}_1,{k}_2) + f({k}_1,{k}_3)
+ f({k}_2,{k}_3)}
{v_{1m}(k_1) v_{1m}(k_1) v_{1n}(k_2) v_{1n}(k_2)+\mathrm{2\ perms.}}.
\eea
The quotient is called $-\frac{6}{5}\f$ and not simply $\f$ because it was
originally defined in terms of the gravitational potential $\gF$ and not $\gz$
as $\gF = \gF_\mathrm{L} + \f \lh \gF_\mathrm{L}^2 
- \langle \gF_\mathrm{L}^2 \rangle \rh$ \cite{Komatsu:2001rj}. 
During recombination (matter domination) the two are 
related by $\gz = - \frac{5}{3} \gF$. Moreover, when computing the bispectrum
divided by the three permutations of the power spectrum squared using this
expression of $\gF$ one obtains $2\f$ due to the two ways the 
two $\gF_\mathrm{L}$ inside the second-order solution can be combined with the 
two linear solutions to create the power spectrum. Together these two effects
explain the factor $-6/5$.\footnote{In \cite{Rigopoulos:2005us} and
earlier papers we used a slightly different definition of $\f$ which was
larger by a factor of -18/5. Here we conform to the definition that is now 
generally accepted in the literature.}

\subsection{\texorpdfstring{ $\delta N$}{DN}-formalism}
\label{dnform}

An alternative formalism to compute $\f$ is the so-called $\delta N$-formalism
\cite{Starobinsky:1986fxa,Sasaki:1995aw,Sasaki:1998ug,Lyth:2004gb,Lyth:2005fi}.
In order to compare our results of the next sections to those obtained
using the $\gd N$ formalism, for those cases where the latter are available,
we give here a brief overview.

The $\gd N$ formalism uses the fact that the adiabatic perturbation $\zeta^1$ 
on large scales is equal to the perturbation of the number of e-folds
$\delta N(t,t_*)$ between an initial flat hypersurface at $t=t_*$, which
is usually taken to be the horizon crossing time, and a final uniform
density hypersurface at $t$. One can then expand the number of e-folds
in terms of the perturbations of the fields and their momenta on the
initial flat hypersurface 
\be \delta N(t,t_*)=\frac{\partial
  N}{\partial\phi^A_*}\delta\phi^A_*+\frac{\partial
  N}{\partial\Pi^A_*}\delta\Pi^A_*
+\frac{1}{2}\frac{\partial^2N}{\partial\phi^A_*\partial\phi^B_*}
\delta\phi_*^A\delta\phi_*^B+\ldots .
\ee 
So instead of integrating the evolution of $\zeta^1$ through
equations (\ref{evol1}) and (\ref{evol2}) one can evaluate the
derivatives of the number of e-folds at horizon crossing and thus
calculate $\zeta^1$.

Because of the computational difficulty associated with the derivatives
with respect to $\Pi^A$, slow roll is assumed at horizon exit so that
the terms involving the momentum of the fields can be ignored. 
This is a crucial assumption for the $\gd N$ formalism.
The final formula then reads 
\be 
\delta N(t,t_*)=\frac{\partial N}{\partial\phi^A_*}\delta\phi^A_*
+\frac{1}{2}\frac{\partial^2N}{\partial\phi^A_*\partial\phi^B_*}
\delta\phi_*^A\delta\phi_*^B,
\ee 
up to second order. From it one finds the following expression for the 
bispectrum:
\be 
\langle \tilde{\zeta}_{\vc{k_1}}^{1}
\tilde{\zeta}_{\vc{k_2}}^{1} \tilde{\zeta}_{\vc{k_3}}^{1}\rangle^{(2)}
=
\frac{1}{2}N_{,A}N_{,B}N_{,CD}\langle \delta\phi^A_{k_1}
\delta\phi^B_{k_2}(\delta\phi^C\star\delta\phi^D)_{k_3}\rangle+\mathrm{perms.},
\ee
where $\star$ denotes a convolution and the average of 
$(\delta\phi \star \delta\phi)$ has been subtracted to avoid divergences. 
$N_{,A}$ denotes the derivative of $N$ with respect to the field $\phi^A_*$. 
Using Wick's theorem this can be rewritten as products 
of two-point correlation functions to yield finally
\be
-\frac{6}{5}f_\mathrm{NL,\delta N} = \frac{N^{,A}N^{,B}N_{,AB}}
{\left(N_{,C}N^{,C}\right)^2}.
\label{dnresult}
\ee 
Notice that this result is momentum independent and local in real space, 
although attempts to generalize to a scale-dependent situation have 
recently been made in \cite{Byrnes:2009pe,Byrnes:2010ft}.
This formula can be used numerically or analytically to calculate $\f$. 
However, for any analytical results and insight one must assume
the slow-roll approximation to hold at all times after horizon exit (see
for example \cite{Vernizzi:2006ve,Choi:2007su,Battefeld:2007en}),
except for the special case of a separable Hubble parameter
\cite{Byrnes:2009qy,Battefeld:2009ym}.

\section{General analytic expression for \texorpdfstring{$\f$}{fNL} 
for two fields}
\label{general}

In this section we will further work out the exact long-wavelength expression 
for $\f$ for the case of two fields, given in (\ref{fNL_start}). 
No slow-roll approximation is used on super-horizon scales in this section. 
In particular this means the formalism can deal with sharp turns in the
field trajectory after horizon crossing during which slow roll temporarily
breaks down. In the first subsection
we restrict ourselves to the case where $k_1=k_2=k_3$ to lighten the 
notation. In the second subsection we show how the result for $\f$ changes 
in the case of arbitrary momenta.

\subsection{Equal momenta}

In the case of equal momenta, equation (\ref{fNL_start})
reduces to
\bea\label{fNL}
-\frac{6}{5}\f =  \frac{-v_{1m}(t) v_{1n}(t)}{\lh v_{1m}(t) v_{1m}(t) \rh^2}
\Bigg\{&&\!\!\int_{-\infty}^t\!\!\!\!\!\d t' G_{1a}(t,t') 
\bA_{abc}(t') v_{bm}(t') v_{cn}(t')
- 2 \get^\perp(t) v_{2m}(t) v_{1n}(t)\nn\\
&&-G_{1a}(t,t_*)M_{abc*} v_{bm}(t_*)v_{cn}(t_*)\Bigg\}.
\eea
We remind the reader that indices $l, m, n$ take the values 1 and 2 (components
in the two-field basis), while indices $a,b,c,\ldots$ take the values 1, 2, 
and 3 (labeling the $\gz^1$, $\gz^2$, and $\gth^2$ components).
To make the expressions a bit shorter, we will drop the
time arguments inside the integrals, but remember that for the Green's
functions the integration variable is the second argument. Using the result
(\ref{timederA}) proved in \ref{appTimeder} we can write $\bA_{ab1}$ as a 
time derivative and do an integration by parts, with the result
	\be\label{fNLint}
	\int_{-\infty}^t\!\!\!\!\!\d t' G_{1a} \bA_{abc}
	v_{bm} v_{cn}
	= 2 \get^\perp v_{2m} v_{1n}
	+ \int_{-\infty}^t\!\!\!\!\! \d t' A_{ab} \frac{\d}{\d t'} 
	[ G_{1a}v_{bm} v_{1n} ]+ 
	\int_{-\infty}^t \!\!\!\!\!\d t' G_{1a} \bA_{ab\bc}
	v_{bm} v_{\bc n},
	\ee
where the index $\bc$ does not take the value 1.
Here we used that the linear solutions $v_{am}$ are zero at $t=-\infty$ 
(by definition), that the Green's function $G_{1a}(t,t) = \gd_{1a}$, and that 
$A_{1b} = -2\get^\perp \gd_{b2}$ (exact).
We see that the first term on the right-hand side exactly cancels 
with the gauge correction (the second term in (\ref{fNL})) that is necessary to 
create a properly gauge-invariant second-order result. 

We start by working out the second term on the right-hand side of
(\ref{fNLint}). We find
\bea
 I && \equiv \int_{-\infty}^t\d t' A_{ab} \frac{\d}{\d t'} 
	\left [ G_{1a} v_{bm} v_{1n} \right ]
         =\gamma_*^2 \int_{-\infty}^t \d t' A_{ab} \frac{\d}{\d t'} 
	\left [ G_{1a} \bv_{bm} \bv_{1n} \gTh(t'-t_*) \right ]\nn\\
	&& = A_{ab*} G_{1a}(t,t_*) v_{bm*} v_{1n*}
	+ \gamma_*^2 \gTh(t-t_*) \int_{t_*}^t \d t' A_{ab} \left [ G_{1d} 
	A_{da} \bv_{bm} \bv_{1n}- G_{1a} A_{bd} \bv_{dm} \bv_{1n}\right.\nn\\ 
	&&\qquad\qquad\qquad\qquad\qquad\qquad\qquad\qquad\qquad\qquad\qquad\left.- G_{1a} \bv_{bm} A_{1d} \bv_{dn} \right ]\nn\\
	&&=A_{ab*} G_{1a}(t,t_*) v_{bm*} v_{1n*}
	- \gamma_*^2 \gTh(t-t_*) \int_{t_*}^t \d t' A_{ab} A_{1d} G_{1a} 
	\bv_{bm} \bv_{dn},
\label{Ainteg}	
\eea
where, as before, a subscript $*$ means that a quantity is evaluated at $t_*$.
Using the explicit form of the matrix $\mx{A}$ (\ref{Amat}) and the solutions 
$v_{am}$ (\ref{vamGreenrel}) this becomes
	\bea
	I & = & \gamma_*^2 \gTh(t-t_*) \gd_{m2} \gd_{n1} 
	\lh -2 \get^\perp_* + \gc_* G_{12}(t,t_*) + A_{32*} G_{13}(t,t_*)
- \gc_* A_{33*} G_{13}(t,t_*) \rh\\
	&& + \gamma_*^2 \gTh(t-t_*) \gd_{m2} \gd_{n2} 
	\int_{t_*}^t \d t' 2\get^\perp \bv_{22} 
	\left [ -2\get^\perp \bv_{22}-G_{12} \bv_{32} 
	+ A_{32} G_{13} \bv_{22} + A_{33} G_{13} \bv_{32} \right ].\non
	\eea
From now on we will drop the overall step function, which just encodes the
obvious condition that $t\geq t_*$.
Realizing that $A_{32}\bv_{22} + A_{33}\bv_{32} = - \frac{\d}{\d t'} \bv_{32}$ 
we can do an integration by parts:
	\bea\label{Ires}
	I = & \gamma_*^2 \gd_{m2} & \Bigg[ \gd_{n1} 
	\lh -2 \get^\perp_* + \gc_* G_{12}(t,t_*) + A_{32*}G_{13}(t,t_*)
	- \gc_* A_{33*} G_{13}(t,t_*) \rh \non\\ 
 && -\delta_{n2}2 \get^\perp_* \gc_*G_{13}(t,t_*)\Bigg]\\
	+\gamma_*^2 & \gd_{m2} \gd_{n2} & \int_{t_*}^t \d t' 2\get^\perp 
\left [ -2\get^\perp (\bv_{22})^2+ 
	\lh - 2 G_{12} + A_{33} G_{13} 
	+ \frac{\dot{\get}^\perp}{\get^\perp} G_{13} \rh \bv_{22} \bv_{32} 
	+ G_{13} (\bv_{32})^2
	\right ].\non
	\eea
To this result we have to add the final term on the right-hand side of 
(\ref{fNLint}). Using the explicit expression for the matrix $\mx{\bA}$ 
and doing
some more integrations by parts this can be worked out further, as can be found
in \ref{derivation}. The final result for $\f$ in the
equal momenta limit is (including also the final term of (\ref{fNL}))
\be\label{fNLresult}
-\frac{6}{5}\f=\frac{-2\bv_{12}^2}{[1+(\bv_{12})^2]^2}
\Bigg( g_{iso}+g_{sr}+g_{int}\Bigg), 
\ee
where
\bea
 g_{iso}&=&(\ge+\get^\parallel) (\bv_{22})^2 + \bv_{22} \bv_{32},
\qquad
g_{sr}=-\frac{\es+\hpas}{2\bv_{12}^2}+\frac{\get^\perp_* \bv_{12}}{2}
-\frac{3}{2}\lh\ge_*+\get^\parallel_*-\gc_*+\frac{\hpes}{\bv_{12}}\rh,\nn\\
 g_{int}&=&- \int_{t_*}^t \d t' \Biggl[ 2 (\get^\perp)^2 (\bv_{22})^2 
	+ (\ge+\get^\parallel) \bv_{22} \bv_{32} + (\bv_{32})^2
- G_{13} \bv_{22} \lh C \bv_{22} + 9 \get^\perp \bv_{32} \rh\Biggr].
\label{gisosrint}
\eea
Here we have defined
\beq
C \equiv 12 \get^\perp \gc - 6 \get^\parallel \get^\perp
	+ 6 (\get^\parallel)^2 \get^\perp + 6 (\get^\perp)^3
- 2 \get^\perp \gx^\parallel- 2 \get^\parallel \gx^\perp
	- \frac{3}{2}(\tW_{211} + \tW_{222}),
\label{Ci}
\eeq
where as before $\tW_{lmn} \equiv (\sqrt{2\ge}/\gk)W_{lmn}/(3H^2)$. 
We should add that although no slow-roll approximation has been used on
super-horizon scales, we did assume slow roll to hold at horizon crossing,
in order to use the analytic linear short-wavelength 
solutions~(\ref{vamGreenrel}) and to remove any dependence on the window
function $\cW$. Observations of the scalar spectral
index seem to indicate that slow roll is a good approximation at horizon 
crossing. In a numerical treatment
we could use the exact numerical solutions instead.

Looking at (\ref{fNLresult}), which is one of the main results of this paper,
we can draw a number of important conclusions.
In the first place there is a part of $\f$, namely the first term in $g_{sr}$,
that survives in the single-field limit. It corresponds to the single-field
non-Gaussianity produced at horizon crossing and comes from the $b_{ia}^{(2)}$
source term. It agrees with the single-field result of Maldacena 
\cite{Maldacena:2002vr} for $\f^{(4)}$.
The rest of the result is proportional to $\bv_{12}$, 
which describes the contribution of the isocurvature mode to the adiabatic
mode. In the single-field case it is identically zero, so that there is no
super-horizon contribution to $\f$ in that case. Moreover, since 
$\gth^1 = 2 \get^\perp \gz^2$, such a contribution only builds up when
$\get^\perp$ is non-zero, i.e.\ when the field trajectory makes a turn in 
field space. We also see that there are three different sorts of terms in
the expression for $\f$. 
The $g_{sr}$ terms are proportional to a slow-roll parameter evaluated at $t_*$ 
and thus are always small because we assume slow roll to hold at 
horizon-crossing. 
Although the terms proportional to $\bv_{12}$ and $1/\bv_{12}$ in $g_{sr}$ are 
time varying, one can easily show that neither $\bv_{12}/(1+\bv_{12}^2)^2$ nor 
$\bv_{12}^3/(1+\bv_{12}^2)^2$ are ever bigger than $0.33$. 
The $g_{iso}$ terms are proportional to
$\bv_{22}$, the pure isocurvature mode. These terms can be big, in particular
during a turn in field space, but in the models that we consider, where the
isocurvature mode has disappeared by the end of inflation, they become
zero again and cannot lead to observable non-Gaussianities. The reason that
we do not consider models with surviving isocurvature modes is that in that
case the evolution after inflation is not clear. In the presence of 
isocurvature modes the adiabatic mode is not necessarily constant (indeed,
that is the source of the non-Gaussianities we are considering here), which
means that the final results at recombination might depend on the details 
of the transition at the end of inflation and of (p)reheating. Hence we will
make sure that in all models we consider the isocurvature modes have
disappeared by the end of inflation, which means in particular that the turn
of the trajectory in field space has to occur a sufficient number of e-folds
before the end of inflation. Note, however, that this is a constraint we
impose voluntarily to simplify the evolution after inflation, it is in no
way a necessary condition for the validity of our formalism during inflation.
Finally, the third type of term in 
(\ref{fNLresult}) is the integral in $g_{int}$. It is
from this integrated effect that any large, persistent non-Gaussianity 
originates.

For completeness we also calculate the power spectrum, which according to 
equation~(\ref{power}) takes the simple form
\be
\mathcal{P}_{\zeta}=\frac{\kappa^2 H_*^2}{8\pi^2\e_*}(1+\bv^2_{12}),
\label{powerspectrum}
\ee
and the spectral index, calculated analytically using 
equations~(\ref{spectral}), (\ref{grstar}) and (\ref{SRder}), 
\bea
\label{spectralindex}
n_{\zeta} - 1 = && \frac{1}{1-\es} \Bigg[-4\e_*-2\hpa_*
+2\frac{\bv_{12}}{1+\bv^2_{12}}\Big(-2\hpe_*+\chi_*\bv_{12} \non\\
 && +G_{13}(t,t_*)\lh -\tilde{W}_{221*}+2\e^2_* +\eta^{\parallel 2}_*
+\eta^{\perp 2}_*  +3\e_*(\hpa_*-\chi_*)-2\hpa_*\chi_*
+\chi_*^2 \rh \Big)\Bigg].
\eea

\subsection{General momenta}
\label{Secgenmom}

We turn now to the more general case where each scale exits the
horizon at a different time $t_{k_i}$, defined by $aH=k_ic/\sqrt{2}$,
where $c \approx 3$ is a constant allowing for some time to pass after horizon
exit so that the long-wavelength approximation is valid (see the discussion
in section~\ref{secGreen}). It is
important to realize that it is not the momentum dependence of the
bispectrum that we are discussing here, but of $\f$.  The momentum
dependence of the local bispectrum is dominated by the momentum
dependence of the power spectrum squared, leading to the well-known
result (see e.g. \cite{Babich:2004gb}) that it peaks on squeezed
triangles where one of the momenta is much smaller than the other
two. Here we are discussing the momentum dependence of $\f$, so one
has divided by the power spectrum squared. This $\f$, often called 
$f_\mathrm{NL}^{(4)}$ in the $\gd N$ literature, is usually assumed to be
momentum-independent. However, as we will show this is not true and
its momentum dependence can lead to relative effects of order $10\%$ even 
within the range of momenta that are observable by Planck.

Assuming $k_1 \geq k_2$, i.e. $t_{k_1} \geq t_{k_2}$, we find that 
(\ref{fkk}) reduces to
\bea
 f(k_1,k_2) &=& -\frac{\gamma^2_{k_1}\gamma^2_{k_2}}{2}
\bv_{1mk_2}(t)\bv_{1nk_1}(t)
\Bigl[\int_{t_{k_1}}^t \d t' G_{1a}(t,t') \left[ \bA_{a\bb\bh}
+ \bA_{a\bh\bb}\right]\bv_{\bb mk_2}(t') \bv_{\bh n k_1}(t')   \non\\
 && 
    -\int_{t_{k_1}}^t \d t' G_{1a}(t,t') \left[ A_{ab}A_{1e}
+ A_{ae}A_{1b}\right]\bv_{bmk_2}(t') \bv_{enk_1}(t')\nn\\
 &&
+ A_{ab}(t_{k_1}) G_{1a}(t,t_{k_1}) \left[ \bv_{bmk_2}(t_{k_1}) 
\delta_{1n}
+ \bv_{1mk_2}(t_{k_1})\delta_{bn} \right]\nn\\
 &&-G_{1a}(t,t_{k1})M_{abc}(t_{k_1})
\left[\bv_{cmk_2}(t_{k_1})\delta_{bn}+\bv_{bmk_2}(t_{k_1})\delta_{cn}\right]
\Bigr],
\eea
where again we have used the result (\ref{timederA}) and have done an
integration by parts that cancels the gauge correction term as in
(\ref{fNLint}). The indices $\bb$ and $\bh$ do not take the value
1. We have introduced the notation $\bv_{i1k_l}\equiv\delta_{i1}$ and
$\bv_{i2k_l}\equiv G_{i2}(t,t_{k_l})-\chi_{k_l}G_{i3}(t,t_{k_l})$,
where $\chi_{k_l}$ is evaluated at $t_{k_l}$.  We notice that due to
the step functions the integral's lower limit corresponds to the time
when both scales have entered the long-wavelength regime, i.e. the
time when the larger $k_1$ (smaller wavelength) exits the horizon. The 
expression has become more complicated as compared to (\ref{fNLint}) and
(\ref{Ainteg}) since the $\bv_{bmk_i}$ refer to a different initial
value depending on the horizon crossing time of each scale $k_i$.
Following the same procedure as in the previous section we find that
\bea\label{fNLgen}
-\frac{6}{5}\f(k_1,k_2,k_3) = \frac{f({k}_1,{k}_2)+f({k}_2,{k}_3)
+f({k}_1,{k}_3)}
{\gamma^2_{k_1}\gamma^2_{k_2}[1+(\bv_{12k_1})^2][1+(\bv_{12k_2})^2]
+\mathrm{2\ perms.}},
\eea
where
\bea
	&&f(k_1,k_2)=-2\gamma^2_{k_1}\gamma^2_{k_2}(\widetilde{v}_{12})^2
\Bigg(g_{iso}({k}_1,{k}_2)+g_{sr}({k}_1,{k}_2)+g_{int}({k}_1,{k}_2)
+g_{k}({k}_1,{k}_2)\Bigg),
\label{fNLresultgenmom}
\eea
with
\bea
 g_{iso}({k}_1,{k}_2)\!&=&\!(\e+\hpa)(\widetilde{v}_{22})^2
+\widetilde{v}_{22}\widetilde{v}_{32},\nn\\	
 g_{sr}({k}_1,{k}_2)\!&=&\!\hpe_{k_1}\Bigg(\frac{G_{22k_1k_2}\bv_{12k_1}}{2}   
-\frac{1}{\bv_{12k_2}}-\frac{G_{22k_1k_2}}{2\bv_{12k_1}}) 
\Bigg)\!+\!\frac{3\chi_{k_2}}{4}G_{33k_1k_2}\nn\\
 	    &&\!\!\!\!\!-\!\frac{3}{2}(\e_{k_1}+\hpa_{k_1})G_{22k_1k_2}
\!+\!\frac{\chi_{k_1}}{4}\Bigg(2\frac{\bv_{12k_1}}{\bv_{12k_2}}+G_{22k_1k_2}\Bigg)-\frac{\e_{k_1}+\hpa_{k_1}}{2(\widetilde{v}_{12})^2},\nn\\
 g_{int}({k}_1,{k}_2)&=&\!-\!\int_{t_{k_1}}^t\!\!\!\!\d t'\Big[2(\hpe)^2
(\widetilde{v}_{22})^2
\!+\!(\e+\hpa)\widetilde{v}_{22}\widetilde{v}_{32}\!+\!(\widetilde{v}_{32})^2
\!-\!G_{13}\widetilde{v}_{22}(C\widetilde{v}_{22}+9\hpe\widetilde{v}_{32})
\Big],\nn\\
 g_{k}({k}_1,{k}_2)&=&\frac{1}{4\bv_{12k_1}}\Big[3G_{13}
(\chi_{k_1}G_{22k_1k_2}\!-\!\chi_{k_2}G_{33k_1k_2})
+G_{32k_1k_2}\left((3\!+\!\e_{k_1}\!+\!2\hpa_{k_1})G_{13}-\bv_{12k_1}
\right)\!\!\Big]\nn\\
	&&\!\!\!\!+\frac{1}{4}G_{12k_1k_2}(-2\hpe_{k_1}+\chi_{k_1}\bv_{12k_1})
+\frac{1}{2}G_{32k_1k_2}(\hpe_{k_1}G_{13}-1)\nn\\
&&\!\!\!\!-\frac{G_{12k_1k_2}}{2\bv_{12k_1}}\lh\e_{k_1}+\hpa_{k_1}+\hpe_{k_1}\bv_{12k_1}\rh
-\frac{1}{2}\lh\e_{k_1}+\hpa_{k_1}-\frac{\chi_{k_1}}{2}\rh G_{12k_1k_2},
\eea
for $k_1\geq k_2\geq k_3$ and $C$ was defined in (\ref{Ci}). We introduced 
the notation
\begin{displaymath}
\begin{array}{l}
(\widetilde{v}_{12})^2\equiv\bv_{12k_1}\bv_{12k_2},\qquad 
(\widetilde{v}_{22})^2\equiv\bv_{22k_1}\bv_{22k_2},\qquad
(\widetilde{v}_{32})^2\equiv\bv_{32k_1}\bv_{32k_2},\\  
\widetilde{v}_{22}\widetilde{v}_{32}\equiv\frac{1}{2}(\bv_{22k_1}\bv_{32k_2}
+\bv_{22k_2}\bv_{32k_1}),
\end{array}
\end{displaymath}
and also $G_{ijk_1k_2}\equiv G_{ij}(t_{k_1},t_{k_2})$, while the
subscript on the slow-roll parameters denotes evaluation at the
relevant time that the scale exits the horizon. The Green's functions
that appear without arguments denote $G(t,t_{k_1})$ outside or
$G(t,t')$ inside the integral.

Although this expression is quite a bit longer than
(\ref{fNLresult}), there are many similarities between the two
results. The whole expression is again proportional to $\bv_{12k_i}$,
except for the single-field horizon-crossing result, so that there is
no super-horizon contribution to $\f$ for the single-field case. 
In the $g_{iso}$ and $g_{sr}$ terms we
recognize the familiar terms of the equal-momenta case, i.e. the
isocurvature contributions proportional to $\bv_{22k_i}$ as well as
the horizon crossing terms now evaluated at $t_{k_1}$ and $t_{k_2}$
(note that for $k_1=k_2$, $G_{iik_1k_2}=1$ identically and we regain
the expressions of (\ref{gisosrint})). The integral has also retained
its form. The rest of the terms, namely those in $g_{k}$, are terms
arising due to the different horizon-crossing times of the scales and
are identically zero for the equal-momenta case $k_1=k_2$ where
$G_{ijk_1k_2}=\delta_{ij}$. All terms inside $g_k$ are proportional
to a slow-roll parameter evaluated at horizon crossing (using the fact
that $G_{13}=G_{12}/3$ up to slow-roll corrections, see (\ref{GsolSR})),
except for the very last term on the second line. 
However, $G_{32k_1k_2}$ is expected to be
quite small: for $k_1=k_2$ it is zero, and for $k_1 \mg k_2$ it becomes 
the linear solution for the isocurvature velocity $\gth^2$ 
(see (\ref{vamGreenrel})). 
Hence we do not expect $g_k$ to give a large 
contribution, which is confirmed numerically. As we will see later on, 
the dominant contribution to the differences between different momentum 
configurations comes from the changes in the other terms.

\section{Slow-roll approximation}
\label{secSlowRoll}

While the exact result for $\f$, equation (\ref{fNLresult}) or 
(\ref{fNLgen}), is an extremely
useful starting point for an exact numerical treatment, the integral cannot
be done analytically. In order to find explicit analytic results that will be 
very useful to gain insight and draw generic conclusions, we need to simplify
the problem by making the slow-roll approximation. In 
subsection~\ref{secSRgenexpr} we further work out (\ref{fNLresult}) under this 
approximation. Even then the integral can only be done analytically for 
certain specific classes of inflationary potentials, which are treated in 
the other subsections.

\subsection{General expressions}
\label{secSRgenexpr}

Considering the slow-roll version of equation (\ref{G22eq}) we find
that $g(t)$ (as defined above equation (\ref{deffY})) 
satisfies
	\beq\label{gSReq}
	\dot{g} + \gc \, g = 0.
	\eeq
We see that $Y(t) \propto \exp(-3t)$ so that $f(t)$ is a rapidly 
decaying solution that can be neglected (see (\ref{deffY}) for definitions). 
After the decaying mode has vanished the solutions for the Green's 
functions simplify to
	\bea\label{GsolSR}
	&&G_{22}(t,t') = \frac{g(t)}{g(t')},\qquad\qquad 
	G_{12}(t,t') = \frac{2}{g(t')} \int_{t'}^t\d\bt\ \hpe(\bt)g(\bt),\nn\\
	&&G_{32}(t,t') = - \gc(t) G_{22}(t,t'),\qquad\qquad 
	G_{x3}(t,t') = \frac{1}{3}G_{x2}(t,t').
	\eea

\subsubsection{Equal momenta}

Using the last two relations in (\ref{GsolSR}) and dropping higher-order terms 
in slow roll, (\ref{gisosrint}) reduces to
\bea
\label{srtot}
 g_{iso}&=&(\e +\hpa-\chi)(\vb_{22})^2,\qquad
g_{sr}=-\frac{\es+\hpas}{2\bv_{12}^2}+\frac{\get^\perp_* \bv_{12}}{2}
-\frac{3}{2}\lh\ge_*+\get^\parallel_*-\gc_*+\frac{\hpes}{\bv_{12}}\rh,\\
 g_{int}&=&\!\!\int_{t_*}^t\!\!\!\!\d t' (\bv_{22})^2 \Bigg[ 2 \get^\perp
\!\lh\! -\! \get^\perp\! +\! \frac{(\ge\!+\!\get^\parallel\!-\!\gc)\gc}
{2\get^\perp}\rh
\!+G_{12} \lh \get^\perp \gc\! -\! 2 \get^\parallel 
\get^\perp\!-\!\frac{1}{2} (\tW_{211} \!+\! \tW_{222}) \rh \Bigg].\non
\eea
Inserting these terms into (\ref{fNLresult}) we find an expression that can 
be considered the final expression for $\f$ in the slow-roll 
approximation, and is the one that will be used in section~\ref{intsection}. 
It also proves useful, however, to rewrite it in a different way using 
integration by parts.

We use the slow-roll version of relation (\ref{Greeneqmot2ftp}),
$2\get^\perp = - \frac{\d}{\d t'} G_{12}(t,t') + \gc G_{12}(t,t')$, to do an
integration by parts, leading to
	\bea\label{Keq}
	g_{int} & = & \bv_{12} \lh -\get^\perp_* + 
	\frac{(\ge_* + \get^\parallel_* - \gc_*)\gc_*}{2 \get^\perp_*} \rh
+ \int_{t_*}^t \d t' G_{12} (\bv_{22})^2 
        \Biggl[ 2 \get^\perp \gc
	- \frac{(\ge+\get^\parallel-\gc)\gc^2}{2\get^\perp} \non\\
	&&- 2 \get^\parallel \get^\perp
	-\frac{1}{2} (\tW_{211} + \tW_{222}) 
	+ \frac{\d}{\d t'} \lh - \get^\perp 
	+ \frac{(\ge+\get^\parallel-\gc)\gc}{2\get^\perp} \rh
	\Biggr].
	\eea
Using the slow-roll version of the relations (\ref{W111W211rel}),
	\beq\label{xiparxiperpSR}
	\gx^\parallel = 3 \ge \get^\parallel + (\get^\parallel)^2
	+ (\get^\perp)^2 - \tW_{111}
	\ \ \ \ \ \ \ \mbox{and}\ \ \ \ \ \ \  
	\gx^\perp = 3 \ge \get^\perp + 2 \get^\parallel \get^\perp
	- \get^\perp \gc - \tW_{211},
	\eeq
as well as the time derivatives of the slow-roll parameters in (\ref{SRder}), 
we can derive that
	\bea
	\frac{\d}{\d t}\lh-\get^\perp +\frac{(\ge+\get^\parallel-\gc)\gc}
{2\get^\perp} \rh
	&=& \frac{1}{2\get^\perp} \Bigg[ -\gc^3 + (\ge+\get^\parallel)\gc^2
- 4 \lh \ge\get^\parallel+(\get^\perp)^2 \rh \gc\\
	&&\!\!\!\!+ 4 \lh \ge^2 \get^\parallel + \ge(\get^\parallel)^2
	- \ge(\get^\perp)^2 + \get^\parallel(\get^\perp)^2 \rh \!-\!(\ge+
\get^\parallel-\gc)\tW_{111}\nn\\
	&&\!\!\!\!+ \lh 2\get^\perp + \frac{(\ge+\get^\parallel-\gc)\gc}
{\get^\perp}\rh \tW_{211} 
	+ (\ge+\get^\parallel-2\gc)\tW_{221}\Bigg].\non
\eea
Inserting this into expression (\ref{Keq}) for $g_{int}$ and including the 
remaining terms in the expression for $\f$ we finally obtain
\bea\label{fNLresultSR}
-\frac{6}{5}\f(t) = && \frac{-2(\bv_{12})^2}{[1+(\bv_{12})^2]^2}
\Biggl\{ (\ge+\get^\parallel-\gc) (\bv_{22})^2
-\frac{\es+\hpas}{2\bv_{12}^2}+\frac{\get^\perp_* \bv_{12}}{2}+\! \frac{(\ge_*\! +\! \get^\parallel_* \!-\! \gc_*)
\gc_*}{2 \get^\perp_*}\, \bv_{12} \non\\ 
&& \qquad\qquad\qquad
-\frac{3}{2}\lh\ge_*+\get^\parallel_*-\gc_*+\frac{\hpes}{\bv_{12}}\rh 
\nn\\
&&+ \int_{t_*}^t \d t' G_{12} (\bv_{22})^2 
\Bigg[2\frac{\ge\get^\parallel}{\get^\perp} 
	\lh\! -\gc\! +\! \ge\! +\! \get^\parallel\! -\!
	\frac{(\get^\perp)^2}{\get^\parallel} \rh\!+\!\frac{1}{2}(\tW_{211}\! 
-\! \tW_{222}\! -\! \frac{\gc}{\get^\perp} \tW_{221})
\nn\\
&& \qquad\qquad\qquad\qquad
- \frac{\ge+\get^\parallel-\gc}{2\get^\perp}
\lh \tW_{111} \!-\! \tW_{221}\! -\! \frac{\gc}{\get^\perp} \tW_{211} \rh
\Bigg] \Biggr\}.
\eea
This is the alternative final result for $\f$ in the slow-roll approximation.

Equation (\ref{fNLresultSR}), as well as (\ref{srtot}), is characterized by the
same features as the result of the exact formalism. We can easily
distinguish the pure isocurvature $\bv_{22}$ term, which we
assume to vanish before the end of inflation in order for the
adiabatic mode to be constant after inflation, as well as the terms evaluated
at the time of horizon crossing, which are expected to be small. Any
remaining non-Gaussianity at recombination has to originate from the integral.
In subsections \ref{secEqPow} and
\ref{intsection} we will further work out the expressions of this section
for the case of certain classes of potentials to gain insight into their 
non-Gaussian properties. But first we look at
the momentum dependence of $\f$ in section~\ref{slowsq}.

\subsubsection{Squeezed limit}
\label{slowsq}

In this section we will calculate the slow-roll expression for $\f$ in the
case where $k \equiv k_3 \ll k_1 = k_2 \equiv k'$, what is usually refered to as 
the squeezed limit. Note that we assume $k_1 = k_2$
for simplicity, to keep the expressions manageable, it is not a necessary
condition.
We start with equation (\ref{fNLgen}) and
follow the procedure of the previous subsection, that is we use the slow-roll
approximations (\ref{GsolSR}) for the Green's functions and drop
higher-order terms in slow roll. Since there are only two relevant
scales the expression simplifies to give
\bea
	-\frac{6}{5}\f&=&\frac{-2\bv_{12k'}/[1+(\bv_{12k'})^2]}
{\gamma^2[1+(\bv_{12k'})^2]+2[1+(\bv_{12k})^2]}
	\Bigg[\gamma^2\bv_{12k'}\!\Bigg(\!g_{iso}(k',k')\!
	+\!g_{sr}(k',k')\!+\!g_{int}(k',k')\!\Bigg) \non\\
	&&+2\bv_{12k}\Bigg(g_{iso}(k',k)\!+\!g_{sr}(k',k)
\!+\!g_{int}(k',k)\!+\!g_{k}(k',k)\!\Bigg)\Bigg],
\label{fNLgeni}
\eea
where $\gamma\equiv\gamma_{k'}/\gamma_{k}$ and
\bea
 g_{iso}(k',k)&=&(\e+\hpa-\chi)\bv_{22k}\bv_{22k'},\nn\\
 g_{sr}(k',k)\!&=&\!\hpe_{k'}\Bigg(\frac{G_{22k'k}\bv_{12k'}}{2}   
-\frac{1}{\bv_{12k}}-\frac{G_{22k'k}}{2\bv_{12k'}}) 
\Bigg)\!+\!\frac{3\chi_{k}}{4}G_{33k'k}\nn\\
 	    &&\!\!\!\!\!-\!\frac{3}{2}(\e_{k'}+\hpa_{k'})G_{22k'k}
\!+\!\frac{\chi_{k'}}{4}\Bigg(2\frac{\bv_{12k'}}{\bv_{12k}}+G_{22k'k}\Bigg)-\frac{\e_{k'}+\hpa_{k'}}{2\bv_{12k}\bv_{12k'}},\nn\\
 g_{int}(k',k)\!&=&\!\!\!\int_{t_{k'}}^t\!\!\!\!\d
t'\bv_{22k}\bv_{22k'}\!\Bigg[ 2 \get^\perp\!\!
	\lh\!\!-\get^\perp\!\!
+\!\frac{(\ge\!+\!\get^\parallel\!-\!\gc)\gc}{2\get^\perp}\!\rh
\!+\!G_{12}\!\lh\! \get^\perp\! \gc\! -\! 2 \get^\parallel
\get^\perp\!\!-\! \frac{1}{2} (\tW_{211}\!
	+\! \tW_{222})\! \rh\!\! \Bigg],\nn\\
 g_{k}({k'},{k})&=&\frac{1}{4}(\chi_{k'}G_{22k'k}\!-\!\chi_{k}G_{33k'k})
+\frac{1}{12}G_{32k'k}(-6+\e_{k'}\!+\!2\hpa_{k'}+2\hpe_{k'}\bv_{12k'})\nn\\
&&\!\!\!\!+G_{12k'k}\lh-\hpe_{k'}-\frac{\e_{k'}+\hpa_{k'}}{2}\lh1+\frac{1}{\bv_{12k'}}\rh+\frac{\chi_{k'}}{4}(1+\bv_{12k'})\rh.
\eea
The first line of (\ref{fNLgeni}), proportional to $\gamma^2$, comes
from the $f(k',k')$ term and it is identical to expression
(\ref{srtot}). The difference is that now it occurs with a weight
$\gamma^2$ compared to the terms originating from $f(k',k)$ that come
with a weight 2. Obviously, in the case of equal momenta where
$G_{ijk_1k_2}=\delta_{ij}$, the expression reduces to equation (\ref{srtot}). 

The $\gamma$ terms can be safely neglected in the squeezed limit because
$\gamma^2$ scales as $e^{-3\Delta t}$, where $\gD t$ is the number of e-folds
between horizon exit of the two scales. If for example the two scales
exit the horizon with a delay $\Delta t\sim 7$, which corresponds to
$k'\sim 1000k$, approximately the resolution of the Planck satellite, we
find that $\gamma^2\sim10^{-9}$.

The functions $\bv_{12}$ and $\bv_{22}$ increase and decrease
respectively (from their initial values 0 and 1) only a little until
the turning of the fields. The later the relevant scale exits the
horizon, the less time there is available for $\bv_{i2}$ to evolve, so
the smaller is the value of $\bv_{12}$ (and the larger for $\bv_{22}$)
during this period.  During the turning of the fields isocurvature
effects turn on. Both $\bv_{12}$ and $\bv_{22}$ vary wildly during
this period. $\bv_{12}$ grows and reaches a constant value afterwards,
while $\bv_{22}$ varies and reaches zero when isocurvature effects
cease.  In the models we studied we found that while during this
period $\bv_{22}$ continues to behave in the same way, i.e. being
larger for the scale that exits last, $\bv_{12}$ changes behaviour and
also becomes larger for the scale that exits last. In the end we
observe that $\f$ in the squeezed limit is smaller than in the
equal-momenta case that was treated in the previous subsection. The
effect is particularly pronounced during the turn of the field
trajectory, mainly due to $g_{iso}$.  As we will show in
section~\ref{secNumQuadrPot}, these effects can reduce the value of
$\f$ during the turn of the field trajectory by $10\%$ on scales that
are within the resolution of Planck.

\subsection{Potentials with equal powers}
\label{secEqPow}
\subsubsection{Quadratic potential}

The quadratic potential has been widely examined in the past and it is
known that it cannot produce large non-Gaussianity (see for example
\cite{Vernizzi:2006ve}).  Here we use our results to analytically
explain why. While the quadratic potential is a special case of the
more general sum potential treated later on, it is still
interesting to discuss it separately in a different way. We start by
deriving the result that for a quadratic two-field potential within
slow roll,
	\beq\label{chirel}
	\gc = \frac{\d}{\d t} \ln \frac{\ge \get^\perp}{\get^\parallel}.
	\eeq
Working out the right-hand side, using (\ref{SRder}), we find
	\beq
	\gc
	= 2\ge + \get^\parallel - \frac{(\get^\perp)^2}{\get^\parallel}
	- \frac{\gx^\parallel}{\get^\parallel} 
	+ \frac{\gx^\perp}{\get^\perp}.
	\eeq
Inserting the relations (\ref{xiparxiperpSR}) (with the third derivatives of the
potential equal to zero, since we have a quadratic potential) this reduces to
	\beq
	\gc = \ge + \get^\parallel - \frac{(\get^\perp)^2}{\get^\parallel}.
	\eeq
It can be checked that this result does indeed satisfy the general equation 
for the time derivative of $\gc$ (\ref{SRder}) within the approximations made,
and the remaining integration constant is fixed by realizing that this result
has the proper limit in the single-field case. This concludes the proof of
(\ref{chirel}).

Since the third-order potential derivatives as well as the first term of the
integral in (\ref{fNLresultSR}) are identically zero, we find that for
a quadratic potential the integral completely vanishes in the slow-roll
approximation and no persistent large non-Gaussianity is produced. 
Numerically we find that even for large mass ratios, when during the turn
of the field trajectory slow roll is broken, the integral is still
approximately zero, see section~\ref{secNumQuadrPot}.

Using this result (\ref{chirel}) for $\gc$ we can also solve (\ref{gSReq}):
	\beq
	g(t) = \frac{\get^\parallel}{\ge \get^\perp},
	\eeq
and hence find that
	\be\label{GsolSRquadr}
	G_{22}(t,t')= \frac{\ge(t') \get^\perp(t')}{\get^\parallel(t')}
	\frac{\get^\parallel(t)}{\ge(t) \get^\perp(t)},\ \ \ \ \ \ \ 
	G_{12}(t,t') = - \frac{\ge(t') \get^\perp(t')}{\get^\parallel(t')}
	\lh\! \frac{1}{\ge(t)} \!+\! 2 t\! -\! \frac{1}{\ge(t')} \!-\! 2t'\!\rh.
	\ee
Note that even though $g(t)$ is a large quantity, of order inverse slow roll, it
is still slowly varying, as we have shown, with its time derivative an order of
slow roll smaller.

\subsubsection{Potentials of the form \texorpdfstring{ $W=\alpha \phi^p+\beta \gs^p$}{}}

For a potential of the form
\be
W(\phi,\gs)=\alpha \phi^p+\beta \gs^q
\label{polynomtype}
\ee
we can work out explicitly the form of the integrand in equation
(\ref{fNLresultSR}). We have to use the slow-roll version of equations
(\ref{fieldeq}) and (\ref{srvar}) to easily find after substitution
that
\be
 g_{int}\!=\!-\!\!\int_{t_*}^t\!\!\frac{\alpha\beta p^4 
(y\!-\!1)\phi^{p-3}\gs^{py-3}\left(y(p y\!-\!1)\phi^2
+(p\!-\!1)\gs^2\right)
\left(\alpha^2\phi^{2 p}\gs^2\!+\!\beta^2 y^2\phi^2\gs^{2 p y}\right)^2\!\!}
{2\kappa^4\left(\alpha\phi^p+\beta\gs^{p y}\right)^4
\left(\alpha(p-1)\phi^p\gs^2-\beta y (py-1)\phi^2\gs^{p y}\right)^2}
\d t',
\ee
where $y\equiv q/p$. 

From this expression we can derive an important
result: for $y=1$, i.e.\ $p=q$, we immediately
see that the integral is zero. This means that no persistent
non-Gaussianity can be produced after horizon exit for potentials
of the form $W(\phi,\gs)=\alpha \phi^p+\beta\gs^p$, at least within
the slow-roll approximation. This generalizes the result for the 
two-field quadratic potential of the previous subsection to any potential 
with two equal powers.

\subsection{Other integrable forms of potentials}
\label{intsection} 

In general, the first step of finding an analytical expression for
the integral $g_{int}$ is to solve the differential equation
(\ref{gSReq}) for $g$ in order to determine the Green's functions.
To do that, one tries to express $\chi$ as a time derivative
of some other quantity. In the slow-roll limit 
\be
\label{etaslow}
\tW_{11} = \ge - \hpa, \qquad\qquad
\tW_{21} = - \hpe,
\ee
so that $\chi$ can be written as
\be
\label{chislow}
\chi=2\e +\tilde{W}_{22}-\tilde{W}_{11}.
\ee
which, as can be checked, cannot be expressed as a derivative of a known 
quantity for a general potential. Thus we are forced to examine special 
classes of potentials.

\subsubsection{Product potentials}

First we consider potentials of the form
\be
W(\phi,\gs)=U(\phi)V(\gs),
\ee
inspired by the analytical study done in 
\cite{Choi:2007su,Byrnes:2008wi}. 
From our point of view, the advantage of these potentials 
is that their mixed second derivative $\tilde{W}_{\phi\gs}$ can be
expressed in terms of the first derivatives to finally give for
the second-order derivatives of the potential in the adiabatic
and isocurvature directions:
\bea
\label{mixed}
 \tilde{W}_{11}&=&\tilde{W}_{\phi\phi}\ef^2+\tilde{W}_{\gs\gs}\ex^2
+4 \epsilon\ef^2 \ex^2,\qquad\qquad 
\tilde{W}_{22}=\tilde{W}_{\phi\phi}\ex^2+\tilde{W}_{\gs\gs}\ef^2
-4\e\ef^2\ex^2,\nn\\
 \tilde{W}_{21}&=&\left(\tilde{W}_{\phi\phi}-\tilde{W}_{\gs\gs}
+2\e(\ex^2-\ef^2)\right)\ef\ex,
\eea
where we used (\ref{e2e1}) to eliminate the unit vector $\vc{e}_{2}$
in terms of $\vc{e}_1$. It is straightforward to show that the second-order
derivatives in the directions of the basis vectors are
related:
\be
\label{w21slow}
\frac{2\e+\tilde{W}_{22}-\tilde{W}_{11}}{\tilde{W}_{21}}=\frac{\ex}{\ef}
-\frac{\ef }{\ex},
\ee
so that only two of them are independent. 
Now we can use (\ref{srvar}) and the above results to write $\chi$ as
\be
\chi=\tilde{W}_{21}\left(\frac{\ex}{\ef}-\frac{\ef}{\ex}\right)
=-\frac{\mathrm{d}}{\mathrm{d}t}\ln\left(\ef\ex\right),
\ee
where the derivatives of the unit vectors are given in (\ref{dere}). 
Hence looking at equation (\ref{gSReq}) we can identify the 
Green's function $g$ to be
\be
g(t)=\ef(t)\ex(t).
\ee
After a few more manipulations the integrand of $G_{12}(t,t')$ 
in (\ref{GsolSR}) takes the form
\be
\hpe(t)g(t)=\frac{1}{4}\frac{\mathrm{d}S}{\mathrm{d}t},
\ee
where $S\equiv\ef^2-\ex^2$, so that the analytical form of the two 
independent linear perturbation solutions in the slow-roll approximation is 
\footnote{Note added: Very recently the same results were obtained 
in \cite{Peterson:2010mv} for the transfer functions $T_\mathcal{{RS}}$ and 
$T_\mathcal{{SS}}$ of product and sum potentials, 
which turn out to coincide with $\bv_{12}$ and $\bv_{22}$.} 
\be
\bv_{12}=\frac{S-S_*}{2 e_{1\phi *}e_{1\gs*}},\qquad\qquad 
\bv_{22}=\frac{e_{1\phi}e_{1\gs}}{e_{1\phi *}e_{1\gs *}}.
\ee

The final step is to write the integrand of $g_{int}$ in (\ref{srtot}) 
in terms of the potential's derivatives and rearrange terms to form
time derivatives. One can prove that then the integrand can be rewritten as
\be
 g_{int}=\frac{1}{1-S_*^2}\int_{t_*}^t \frac{\d}{\d t'}
\Bigg[\Big(S(t)-S(t')\Big)
\Big(\tilde{W}_{\sigma\sigma}(t')\ef^2(t')-\tilde{W}_{\phi\phi}(t')
e^2_{1\sigma}(t')\Big)\Bigg]\d t'.
\ee
After performing the integration and adding the rest
of the terms we find that
\be
-\frac{6}{5}\f=\frac{2(S-S_*)^2(S_*^2-1)}{(1+S^2-2SS_*)^2}
\left(g_{iso}+g_{sr}+g_{int}\right),
\label{fpro}
\ee
where now
\bea
 g_{iso}&=&
\frac{S^2-1}{S_*^2-1}\left(\e+\hpa-\chi\right),
\nn\\ 
 g_{sr}&=&
-\frac{1}{2(S-S_*)}\Bigg[\lh\es+\hpas\rh\frac{1+3S(S-2S_*)+2S_*^2}
{S-S_*}-\chi_*\frac{-3+S^2+4SS_*-2S_*^2}{2S_*}\Bigg],\nn\\
 g_{int}&=&
-\frac{S_*(S-S_*)}{S_*^2-1}
\left(\es+\hpas-\chi_*\frac{S_*^2+1}{2S_*^2}\right).
\eea
Comparing
to the results of \cite{Choi:2007su,Byrnes:2008wi} we find complete 
agreement. 

Looking at the result for $\f$ for the product potential we can draw a number
of conclusions. The only time-dependent
slow-roll parameters appear in $g_{iso}$. These terms and consequently 
$\f$ can vary significantly during a turn of the field trajectory but, as 
explained before, in the models we consider isocurvature modes have
disappeared by the end of inflation so that the adiabatic mode will be 
constant after inflation, which means $g_{iso}$ will disappear again and
cannot give any persistent non-Gaussianity.
The rest of the terms involve slow-roll parameters evaluated at horizon
crossing, which are small.
Hence we conclude that any large non-Gaussianity will have to come from
the denominator becoming very small (since $|S| \leq 1$ the numerator cannot
become large) to compensate for the small slow-roll parameters at horizon 
crossing. We see that this can only happen when $S,S_* \rightarrow \pm 1$.
In the remainder of this section we will study the two different cases
that satisfy this condition: a $90^\circ$ turn in the field trajectory
($S=-S_*$), or the same field dominating both at the beginning and at the end
($S=S_*$).

First we study the case where the field trajectory makes a $90^\circ$ turn.
The field $\phi$ is dominant right after horizon crossing, which means 
$|\exs|\ll 1$, $|\efs| \approx 1$ and hence $S_*\rightarrow 1$. 
Later on occurs a turn in the field trajectory and afterwards $\gs$ leads 
inflation, so that $|\ef|\ll 1$, $|\ex|\approx 1$ and $S\rightarrow-1$.  
Then we find that both $g_{sr}$ and $g_{iso}$ go to zero, which means 
in particular that we satisfy the condition on the disappearance of the 
isocurvature mode that allows us to directly extrapolate the results at 
the end of inflation to the time of recombination. 
The non-zero term comes as expected from $g_{int}$
and it is given by:
\be
-\frac{6}{5}\f = \ge_* + \hpa_* - \gc_* = - \tW_{\gs\gs*}, 
\ee
since 
$\tW_{\gs\gs*}=\tW_{22*}=\gc_*-\ge_*-\hpa_*$.
Hence we see that for any product potential where the field trajectory
makes a $90^\circ$ turn no significant non-Gaussianity will be produced,
at least within the slow-roll assumptions used to derive this analytic
result.

Next we look at the opposite limit, where one of the fields, $\phi$,
is dominant both at horizon crossing and at the end of inflation.
This means $|\exs|\ll 1$, $|\efs| \approx 1$ and $|\ex|\ll 1$, $|\ef|
\approx 1$, so that $S_*\rightarrow 1$ and $S\rightarrow 1$. This
includes the case where we have a perfectly straight field trajectory,
i.e. an effectively single-field situation, where obviously no
super-horizon non-Gaussianity is produced. However, we find that even
more generally in this limit the contributions from $g_{iso}$ and
$g_{int}$ go to zero and we are left with only the single-field result
from $g_{sr}$:
\be
-\frac{6}{5}\f =\ge_* + \hpa_*.
\ee
Hence no significant non-Gaussianity is produced in this limit.

We conclude that if we impose the condition of the disappearance of the
isocurvature mode by the end of inflation, to simplify the evolution
afterwards, the product potential can never give large non-Gaussianity, 
at least within the slow-roll approximation.

\subsubsection{Potentials of the form \texorpdfstring{ $W(\phi,\gs)=(U(\phi)+V(\gs))^{\nu}$}{(U+V)n}}
\label{uvn}

Next we consider potentials of the form
\be
W(\phi,\gs)=(U(\phi)+V(\gs))^{\nu},
\ee
which, to our knowledge, have not been worked out before for general 
$\gn$.\footnote{Note added: While we were doing the final editing of our manuscript,
a paper \cite{Meyers:2010rg} appeared on the arXiv where the authors
studied this type of potential using the $\gd N$ formalism. Their result
for $\f$ agrees with ours.}
While of course not the most general two-field potential, 
it can accommodate potentials with coupling terms of the form 
$\alpha^2\phi^2+\beta^2\gs^2+2\alpha\beta\phi\gs$ or higher-order 
combinations.
Note that in the case of $\gn=1$ the potential becomes the simple sum potential,
which has been studied before \cite{Vernizzi:2006ve,Byrnes:2008wi}.

Just as for the product potential, we find that mixed second derivatives 
of the potential can be expressed in terms of the other derivatives:
\bea
&&\tilde{W}_{11}\!=\!\tilde{W}_{\phi\phi}\ef^2+\tilde{W}_{\gs\gs}\ex^2
+\frac{4\e(\nu -1)\ef^2\ex^2}{\nu },\ \ 
\tilde{W}_{22}\!=\!\tilde{W}_{\gs\gs}\ef^2+\tilde{W}_{\phi\phi}\ex^2
-\frac{4\e(\nu -1)\ef^2\ex^2}{\nu },\nn\\
&&\tilde{W}_{21}\!=\!\left(\tilde{W}_{\phi\phi}-\tilde{W}_{\gs\gs}
\right)\ef\ex+\frac{2\e(\nu -1)\ef\ex(\ex^2-\ef^2)}{\nu }.
\eea
Again there are only two independent second derivatives of the potential
in our basis:
\be
\label{w21sum}
\tilde{W}_{21}=\left(\tilde{W}_{22}-\tilde{W}_{11}+\frac{2\e(\nu-1)}{\nu}\right)\frac
{\ex\ef} {\ex^2-\ef^2}.
\ee
Following the procedure of the previous section we rewrite $\chi$ as
\be
\chi=\frac{2\e}{\nu} +\tilde{W}_{21}\left(\frac{\ex}{\ef}-\frac{\ef}{\ex}\right)
=-\frac{\mathrm{d}}{\mathrm{d}t} \ln\left(H^{2/\nu}\ef\ex\right),
\ee
and then find an analytical expression for $g$,
\be
g(t)=H^{2/\nu}(t)\ef(t)\ex(t).
\ee
The integrand of $G_{12}(t,t')$ is now written as
\be
\hpe(t)g(t)=\frac{1}{2}\left(\frac{\kappa^2}{3}\right)^{1/\nu}\frac
{\mathrm{d}
Z}{\mathrm{d}t},
\ee
where $Z\equiv V\ef^2-U\ex^2$. Finally we find that
\be
\bv_{12}=\frac{Z-Z_*}{W_*^{1/\nu} e_{1\phi *}e_{1\gs*}},\qquad\qquad
\bv_{22}=\frac{W^{1/\nu} e_{1\phi}e_{1\gs}}{W_*^{1/\nu}
e_{1\phi *}e_{1\gs *}}
\label{G12}.
\ee

Rewriting the integrand of $g_{int}$ in terms of the potential's derivatives yields after a few manipulations
\bea
 g_{int}=\frac{W_*^{-2/\nu}}{\efs^2\exs^2}&&\int_{t_*}^t\!\!\!\d t'
\Bigg\{\frac{\d}{\d t'}\Big[\frac{W^{2/\nu}(t')\e(t')\ef^2(t')
e_{1\sigma}^2(t')}{\nu}\Big]\nn\\ 
 &&+\frac{\d}{\d t'}\Big[W^{1/\nu}(t')\frac{Z(t)-Z(t')}{2}
\Big(\tilde{W}_{\sigma\sigma}(t')\ef^2(t')-\tilde{W}_{\phi\phi}(t')
e^2_{1\sigma}(t')\Big)\Big]\Bigg\}
\eea
and adding the rest of the terms results in
\be
-\frac{6}{5}\f\!=\!-\frac{2W_*^{2/\nu}(Z-Z_*)^2\efs^2\exs^2}
{\left(\exs^2(Z+U_*)^2\!+\!\efs^2(Z-V_*)^2\right)^2}
\Bigg(g_{iso}+g_{sr}+g_{int}\Bigg),
\label{nuv}
\ee
where
\bea
 g_{iso}&=&
\left(\frac{W^{1/\nu}\ef\ex}{W_*^{1/\nu}\efs\exs}\right)^2
\Bigg(\e+\hpa-\chi\Big),\nn\\
 g_{sr}&=&
-\frac{3}{2}(\es+\hpas-\chi_*)+\frac{Z-Z_*}{W_*^{1/\nu}(\efs^2-\exs^2)}
(-\frac{\es}{\nu}+\frac{\chi_*}{2})
\Bigg[1-\frac{3W_*^{2/\nu}\efs^2\exs^2}{(Z-Z_*)^2}\Bigg]\nn\\
 &&\!\!\!\!-(\es+\hpas)\frac{W_*^{2/\nu}\efs^2\exs^2}{2(Z-Z_*)^2}\nn\\
 g_{int}&=&
\frac{Z-Z_*}{2W_*^{1/\nu}}\left(\frac{1}{\exs^2}-\frac{1}{\efs^2}\right)
\Bigg(\es+\hpas-\frac{\chi_*}{2}\Big(1+\frac{1}{(\efs^2-\exs^2)^2}\Big)
\Bigg)\nn\\
&&\!\!\!\!+\frac{\e}{\nu}\left(\frac{W^{1/\nu}\ef\ex}{W_*^{1/\nu}\efs\exs}
\right)^2-\frac{\es}{\nu}
\Big(1-\frac{2(Z-Z_*)}{W_*^{1/\nu}(\efs^2-\exs^2)}\Big).
\label{nuvt}
\eea
Note that the first term on the second line of $g_{int}$ is also
related to the pure isocurvature mode, but we have not incorporated it
in $g_{iso}$ in order to remind the reader that it originates from the
integral. 

As in the case of the product potential we will study two limiting cases,
to get some insight into the behaviour of $\f$. First is the limit where
the field trajectory makes a $90^\circ$ turn. We assume that $\phi$ 
dominates inflation at horizon exit, that is $|\exs| \ll 1$,
$|\efs| \approx 1$ and $Z_* \rightarrow V_*$. At late times, after the 
turn of the field trajectory, the second field $\gs$ is dominant and the 
contribution of $\phi$ is negligible, so that $|\ef|\ll 1$,
$|\ex| \approx 1$ and $Z \rightarrow -U$. 
Then we find that $g_{sr}$ and $g_{iso}$ go to zero, while the remaining
contribution to $\f$ comes from $g_{int}$, as expected,
\be
-\frac{6}{5}\f=-\frac{U_*+V_*}{U+V_*}\tW_{\gs\gs *}
= \frac{U_*+V_*}{U+V_*} \lh \ge_* + \hpa_* - \gc_* \rh. 
\label{lim}
\ee
So we see that we need a significant decrease in $U$ between horizon crossing
and the end of inflation, as well as a relatively small value of $V_*$, 
to get a large $\f$. Of course we cannot increase
$U_*/U$ too much without breaking slow roll.
In section~\ref{secNewPot} we investigate numerically the properties of a 
model with a sum potential and confirm the validity of the above limit.

In the opposite limit $\gf$ dominates both at horizon crossing
and at the end of inflation, i.e. $|\exs|\ll 1$, $|\efs| \approx 1$ and 
$|\ex|\ll 1$, $|\ef| \approx 1$ so that $Z_* \rightarrow V_*$ and
$Z \rightarrow V$. Then the expression reduces to
\be
-\frac{6}{5}\f = -\frac{U_*+V_*}{V-V_*} \lh \ge_* + \hpa_* - \gc_* \rh,
\ee
which comes from $g_{int}$. Note that we have assumed here that $V \neq V_*$. 
In the (effectively)
single-field case this is not valid; in that case we find that $g_{int}$
and $g_{iso}$ are zero and $g_{sr}$ goes to the single-field result,
$\ge_* + \hpa_*$. 
We remark that in this limit $g_{iso}$ is zero, so that the
adiabatic mode is conserved after inflation.
In order to make $\f$ large, one might be tempted to take $V$ close to
$V_*$. However, that means $\sigma$ does not evolve and we are in an
effectively single-field situation, where the above limit is not
valid. Instead the situation is somewhat similar to the previous
limit: we need a large value of $U_*$ and relatively small values of
$V_*$ and $V$ to overcome the small values of the slow-roll parameters
at horizon crossing. It might not be simple to satisfy these
conditions together with the requirements of this limiting case that
$\phi$ dominates both at horizon crossing and at the end of inflation,
with a period of $\sigma$ domination in between; we did not further
study those types of models.

As a final remark we point out that the power $\nu$ of the potential
does not appear explicitly in the limits for $\f$. Of course its value
will play a role in determining the field trajectory and the values
of the slow-roll parameters, but that is only a relatively small effect.
We have verified this result numerically for several values of the power
$\nu$ of sum potentials of the form (\ref{polynomtype}).

\section{Numerical results}
\label{numerical}

The formalism we have developed so far provides a tool to calculate
the exact amount of non-Gaussianity produced during inflation driven
by a general two-field potential, beyond the slow-roll
approximation. While we assumed slow roll in the previous section,
in order to derive analytical results, we return here to the
exact formalism for a numerical treatment.
In the following subsections we investigate the
properties of the quadratic potential as well as a potential of the
sum type that can produce an $\f$ of the order of a few, and compare our
results to those of the $\delta N$-formalism.

\subsection{Comparison with \texorpdfstring{ $\delta N$}{DN} for the quadratic 
potential}
\label{secNumQuadrPot}

We investigate the quadratic potential
\be
W=\frac{1}{2}m_{\phi}^2\phi^2+\frac{1}{2}m_{\gs}^2\gs^2
\label{model}
\ee
choosing our parameters as follows: $m_{\phi}/m_\gs=20$, 
$m_{\gs}=10^{-5}\kappa^{-1}$ and the initial conditions 
$\phi_0=\gs_0=13\kappa^{-1}$ at $t=0$ for a total of about 85
e-folds of inflation. From now on we will denote the heavy field as $\phi$. 
We choose to present this particular mass ratio because the fields oscillate 
wildly during the turn and slow roll is badly broken, so that it provides a 
serious check both of our formalism and the $\gd N$ one. Of course we have
also run tests with smaller mass ratios when slow roll is unbroken and verified
our analytical slow-roll results.

\begin{figure}
\begin{tabular}{cc}
\includegraphics[scale=0.8]{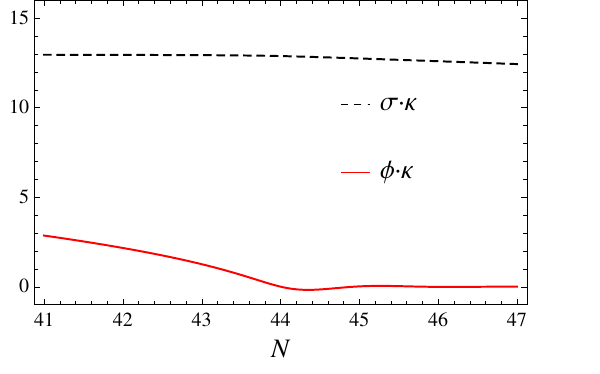}
&\includegraphics[scale=0.75]{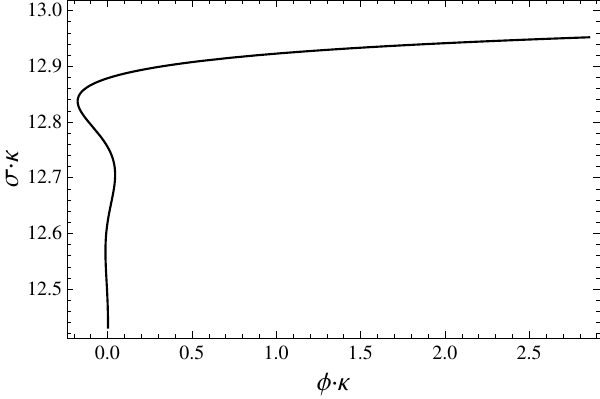}
\end{tabular}
\caption{The time evolution of the fields (left) and the field trajectory
(right) during the period of the turn of the field trajectory,
for the model (\ref{model}) with initial conditions 
$\phi_0=\gs_0=13\kappa^{-1}$ and mass ratio $m_{\phi}/m_{\gs}=20$.}
\label{fig1}
\end{figure}

\begin{figure}

\begin{tabular}{cc}
\includegraphics[scale=0.8]{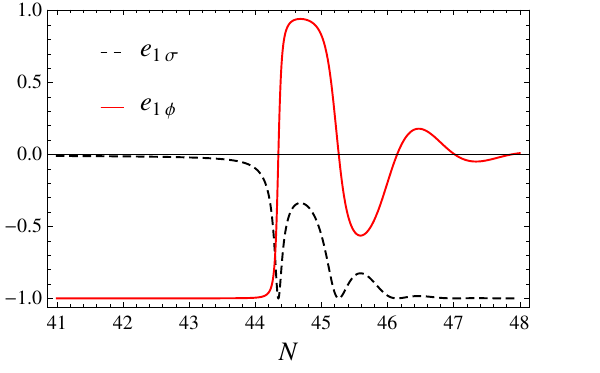}
&\includegraphics[scale=0.8]{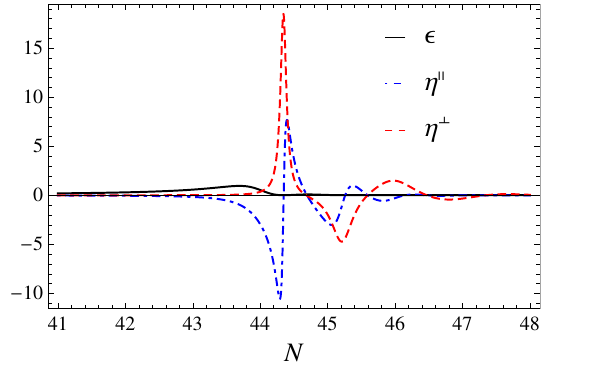}
\end{tabular}
\caption{The unit vectors (left) and the slow-roll parameters 
$\e,\hpa$ and $\hpe$ (right) as a function of time during the turn of the
field trajectory, for the same model as in figure~\ref{fig1}.}
\label{fig2}
\end{figure}

We solve the field equations (\ref{fieldeq}) numerically and in
figures~\ref{fig1} and \ref{fig2} we plot the values of the fields,
the unit vectors, and the slow-roll parameters as a function of time
during the range of e-folds where the heavy field $\phi$ is approaching zero
and starts oscillating. In
the beginning of inflation $\phi$ dominates the expansion while
rolling down its potential and about 40 e-folds after the initial
time $t=0$ it starts oscillating around the minimum of its
potential. The heavier $\phi$ is, the more persistent are the damped
oscillations. During the period of oscillations the unit vectors, as
well as the slow-roll parameters $\e$, $\hpa$, and $\hpe$, oscillate
too. For $m_{\phi}/m_\gs=20$ the maxima of
the slow-roll parameters are much larger than unity and slow roll is
temporarily broken. 
During these oscillations the light field $\gs$ starts driving 
inflation and rolls down its potential until it also reaches its minimum and
starts oscillating. We take the end of inflation when $\e=1$ during this
second period of oscillations.  The situation is
similar to the limiting case we studied in section \ref{uvn} with
$|\exs|\ll1$ and $|\ef|\ll1$.

\begin{figure}
\begin{center}
\begin{tabular}{ll}
\includegraphics[scale=1]{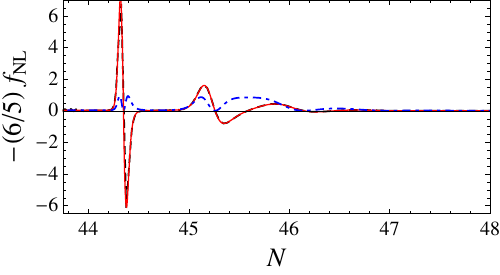}
&\includegraphics[scale=1]{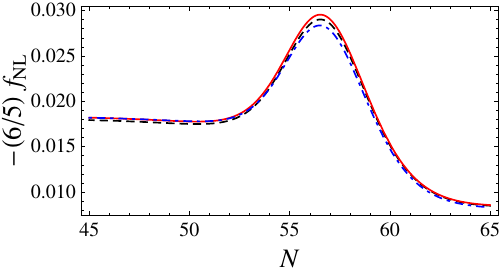}
\end{tabular}
\caption{We plot the $\f$ parameter for the model (\ref{model}) with initial 
conditions $\phi_0=\gs_0=13\kappa^{-1}$ and mass ratio $m_{\phi}/m_{\gs}=20$ 
(left) and $m_{\phi}/m_{\gs}=4$ (right).
The red line is the exact numerical result, while the blue dot-dashed line 
shows 
the slow-roll analytical approximation (but using the exact background).
We also show the numerical $\delta N$ result as the black dashed line, which
lies practically on top of our red result.}
\label{fig3}
\end{center}
\end{figure}

In figure \ref{fig3} we plot the $\f$ parameter as calculated in
our formalism both the numerical exact version (\ref{fNLresult}) 
and the slow-roll 
analytical approximation (\ref{fNLresultSR}) (but using the 
exact background), as well as the result computed numerically
in the context of the $\delta N$-formalism.  The horizon-crossing
time is defined as 60 e-folds before the end of inflation. We do not
expect any large non-Gaussianity to be produced in this model,
since we have shown that the integral of (\ref{fNLresultSR}) is equal
to zero in the slow-roll approximation. The final
value of $\f$ calculated in all three cases is $\mathcal{O}(10^{-2})$. 
Our results coincide completely with those of
the $\delta N$-formalism, thus reinforcing the validity of both
formalisms.
We also show $\f$ for a much smaller mass ratio, $m_\gf/m_\gs = 4$,
where slow roll remains valid throughout the turn of the field trajectory,
verifying our analytical slow-roll result.

The peak of the $\f$ parameter during the turning of the fields is due
to the isocurvature terms $g_{iso}$ in the slow-roll analytical formula.
As expected this effect is transient and disappears
when the isocurvature mode $\bv_{22}$ has been fully converted to the 
adiabatic one. There is no surviving isocurvature mode in this model.
The higher is the mass ratio, the larger is the magnitude of
the peak as a consequence of the more violent oscillations.

\begin{figure}
\begin{center}
\begin{tabular}{cc}
\includegraphics[scale=0.9]{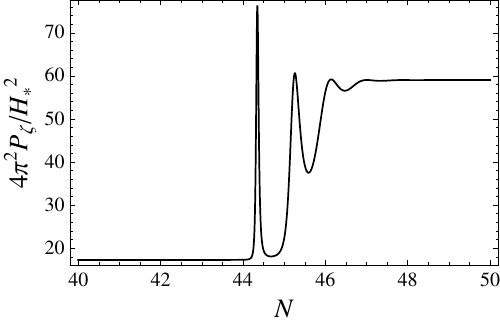}
&\includegraphics[scale=0.9]{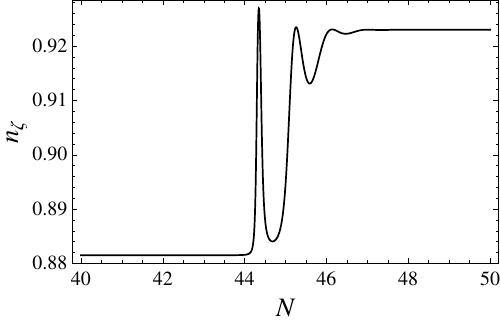}
\end{tabular}
\caption{The exact numerical power spectrum (left) and the spectral 
index (right) for the same model as in figure~\ref{fig1}.}
\label{fig4}
\end{center}
\end{figure}

For completeness, we plot in figure~\ref{fig4} the power spectrum 
(\ref{powerspectrum}) and the spectral index (\ref{spectralindex}) 
of this model. We see there is a jump in both of them
during the oscillatory period of the heavy field, but afterwards they
become constant again.

\begin{figure}
\begin{center}
\begin{tabular}{cc}
\includegraphics[scale=0.9]{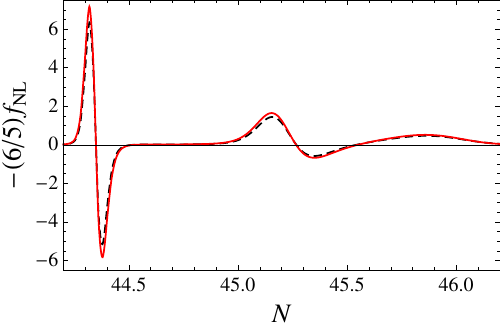}
&\includegraphics[scale=0.9]{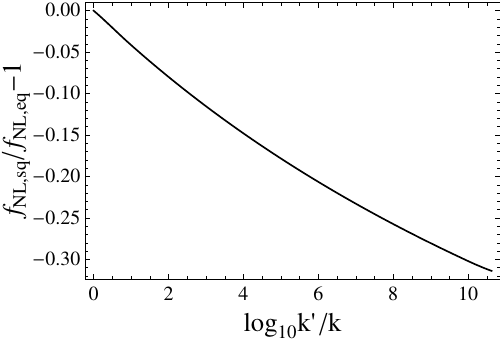}
\end{tabular}
\caption{In the first plot we depict the $\f$ parameter for the equal 
momenta limit in red and the squeezed limit 
result for $k'/k=1000$ in dashed black. In the second plot we show the 
dependence of the discrepancy of the first peak value of $\f$ in the two 
limits on the ratio $k'/k$. We used the same model as in figure~\ref{fig1}
and $t_{k'}=25$ (60 e-folds before the end of inflation).}
\label{fig5}
\end{center}
\end{figure}

Finally in figure~\ref{fig5} we plot the exact numerical $\f$ in
the squeezed limit and in the equal momenta limit. As mentioned in 
section~\ref{slowsq}, 
we see that the $\f$ parameter in the squeezed limit is smaller than 
in the equal-momenta one. 
From figure \ref{fig5} we can see that for $k'=1000k$
(roughly corresponding to the Planck resolution) the peak value of
$\f$ is more than $10\%$ smaller than for $k'=k$, for this particular model.

\subsection{A simple model producing large non-Gaussianity}
\label{secNewPot}

In this section we introduce a model that produces an $\f$ of the order of 
a few, which is two orders of magnitude larger than the single-field 
slow-roll result. So in that sense we can call it large. From the point of 
view of observations with the Planck satellite it is probably still a little 
bit too small, but we have taken this particular model to be able to make
the connection with our analytical results.

The $\f$ limit (\ref{lim}) that we calculated in section \ref{uvn} can
be simplified for the sum potential ($\nu=1$) to give
\be
-\frac{6}{5}\f=-\frac{V_{\gs\gs *}}{\kappa^2(U+V_*)},
\ee
where we used the definition of $\tilde{W}_{mn}$ and the slow-roll
version of (\ref{fieldeq}) for $H$.  We can easily infer that in order
to obtain a large value for $\f$, the heavy field $\phi$ should end up
with a small value at the end of inflation, while $\gs$ should obey a
potential characterized by a large second derivative and a small value
at horizon crossing. Such properties can be accommodated by
a potential of the form
\bea
\label{newpot}
U(\phi)&=&a_2\phi^2,\nn\\
V(\gs)&=&b_0-b_2\gs^2+b_4\gs^4,
\eea
with $b_0=b_2^2/(4b_4)$ so that the minimum of the potential has $W=U+V=0$.

\begin{figure}
\begin{center}
\begin{tabular}{cc}
\includegraphics[scale=0.75]{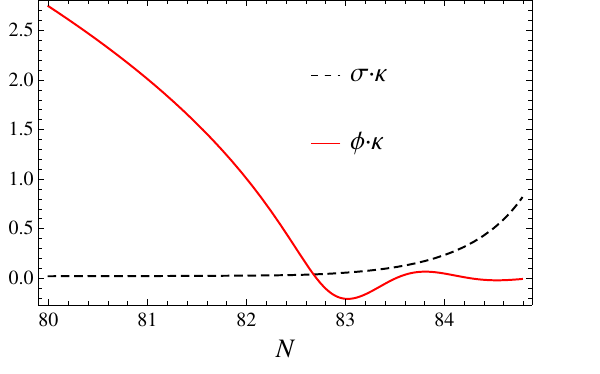}
&\includegraphics[scale=0.68]{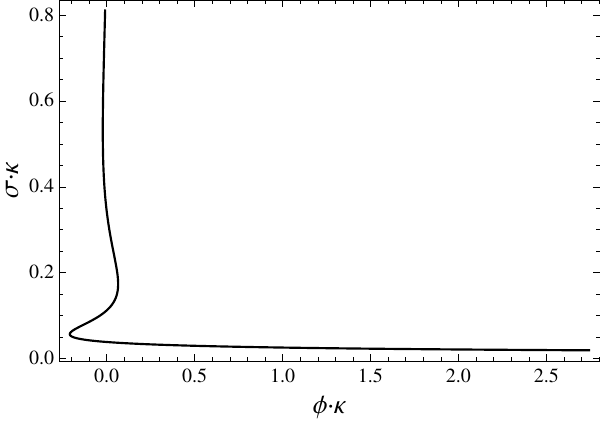}
\end{tabular}
\caption{The time evolution of the fields (left) and the field trajectory
(right), for the model (\ref{newpot}) with initial conditions 
$\phi_0=18\kappa^{-1},\gs_0=0.01\kappa^{-1}$ and parameters 
$a_2=20\kappa^{-2},b_2=7\kappa^{-2}$ and $b_4=2$. Only the time interval
during the turn of the field trajectory is shown.}
\label{fig6}
\end{center}
\end{figure}

\begin{figure}
\begin{center}
\begin{tabular}{cc}
\includegraphics[scale=0.8]{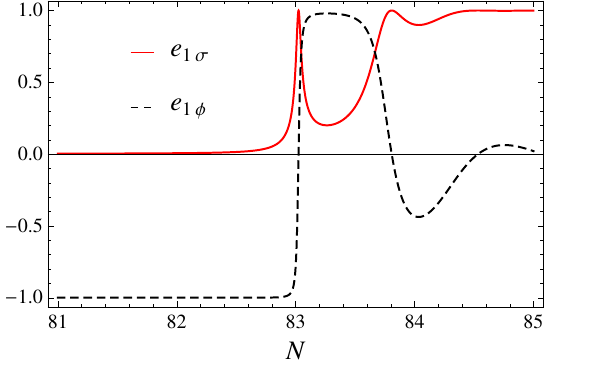}
&\includegraphics[scale=0.82]{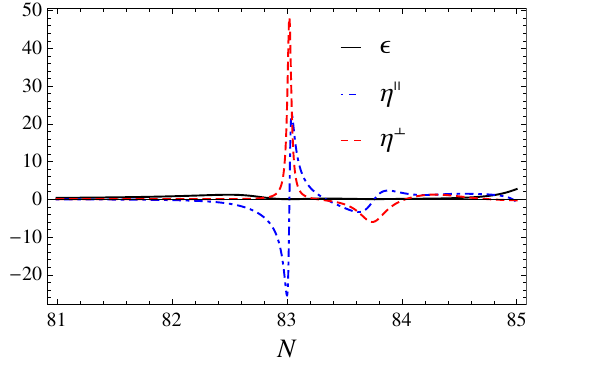}
\end{tabular}
\caption{The unit vectors (left) and the slow-roll parameters 
$\e,\hpa$ and $\hpe$ (right) as a function of time during the turn of the
field trajectory, for the same model as in figure~\ref{fig6}.}
\label{fig7}
\end{center}
\end{figure}

To illustrate the above we investigate a model with
$a_2=20\kappa^{-2},b_2=7\kappa^{-2}$, $b_4=2$, and initial
conditions $\phi_0=18\kappa^{-1}$ and $\gs_0=0.01\kappa^{-1}$, so
that the light field is standing on the local maximum of its
potential, for a total amount of 85 e-folds of inflation. 
This type of effective potential might be realized in the
early universe during second-order phase transitions.  We solve
the field equations (\ref{fieldeq}) numerically and in figures
\ref{fig6} and \ref{fig7} we plot the evolution of the fields and the 
unit vectors, as well as the slow-roll parameters. The situation is
qualitatively the same as in the case of the quadratic potential: in
the beginning $\phi$ dominates inflation while rolling down its
potential, then there is a period of violent oscillations around $\gf=0$, 
and $\gs$ takes over and starts rolling down towards the minimum 
of its potential.

The behaviour of the unit vectors is that of the limiting case we
studied in section \ref{uvn}, that is $|\exs|\ll1$ and $|\ef|\ll1$. We will try 
to obtain an analytical estimate for the magnitude of $\f$. The final value of 
$\f$ is reached when the fields have rolled down to
their minima, that is at $\phi=0$ and $\gs=\sqrt{b_2/(2b_4)}$ for a positive
initial condition for $\gs$. Then $\f$ becomes
\be
-\frac{6}{5}\f=
8\frac{\omega(1-6\omega\gs_*^2)}{\kappa^2(1-2\omega\gs_*^2)^2},
\ee
where $\omega\equiv b_4/b_2$. Note that within our approximation $\f$
depends only on the value of $\sigma_*$ at horizon crossing once we
have fixed the ratio $\omega$. Since the turning of the fields occurs
only a few e-folds before the end of inflation, we will explicitly
assume $W\simeq U$ is a good approximation for nearly all the period
of inflation.  Then we can solve the field equations in the slow-roll
approximation to find
\be
 \phi(t)=\phi_0\sqrt{1-\frac{4t}{\kappa^2\phi_0^2}},\qquad\qquad
\gs(t)=\Bigg[2\omega-\left(2\omega-\frac{1}{\gs_0^2}\right)
\left(1-\frac{4t}{\kappa^2\phi_0^2}\right)^{r}\Bigg]^{-1/2},
\ee
where $r\equiv b_2/a_2$.

The time of horizon crossing $t_*=t_{fin}-60$ can be approximately
found from the final time $t_{fin}\simeq\kappa^2\phi_0^2/4$ and thus
we calculate the values of the fields at horizon exit as functions of
the initial conditions $\phi_0$ and $\gs_0$.  Using these results in
$\f$ we find 
\be
-\frac{6}{5}\f=-\frac{8\omega}{\kappa^2\left(1-2\omega\gs_0^2\right)^2}
\Bigg[2\left(-1+\tilde{\phi}_0^{2r}\right)\omega\gs_0^2+1\Bigg]
\Bigg[2\left(1+2\tilde{\phi}_0^{2r}\right)\omega\gs_0^2-1\Bigg], 
\ee
where $\tilde{\phi}_0=\phi_0/(2\sqrt{60}/\kappa)=\phi_0/\phi_*$.

We now check the dependence of the above expression on the initial 
condition $\gs_0$. Since we assumed that $|\exs|\ll1$ 
and $W\simeq U$ we examine the case $\gs_0\ll 1$ where 
\be
-\frac{6}{5}\f=\frac{8\omega}{\kappa^2}(1-2\omega\gs_0^2\tilde{\phi}_0^{2r})
\label{an}
\ee 
up to second order with respect to $\gs_0$. The parameter $\f$ becomes 
maximal if $\omega=\tilde{\phi}_0^{-2r}/(4\gs_0^2)$ and its 
value is then
\be
-\frac{6}{5}\f=\frac{\tilde{\phi}_0^{-2r}}{\kappa^2\gs_0^2}.
\ee
Since $\tilde{\phi}_0>1$, the smaller the ratio $r$ and the smaller the 
initial value of the field $\gs$, the higher is the value of $\f$.

Nevertheless one has to assure that the turn of the field trajectory
does not occur too late (too close to the end of inflation), so that the
isocurvature mode will have had the time to disappear before the end of
inflation (so that we can directly extrapolate the results at the end of
inflation to the time of recombination and do not have to take further
evolutionary effects into account) and the oscillations of the heavy field 
do not coincide with those of the light field. 
The higher is the ratio $\omega$, the larger is $\f$, but
then the minimum of the potential approaches $\gs_0$ and consequently
there is less time available for $\bv_{12}$ and thus for the
adiabatic perturbation to become constant. This turns out to be a
non-trivial requirement: although we do not claim to have scanned the
whole parameter space of the model, we could not find parameter
values that passed the above test and at the same time yielded a very
large $\f$. The values we have chosen to work with in this paper
respect the above condition and using expression (\ref{an}) we expect
to find $-(6/5)\f\sim 2$. 

If one were to take $b_4=5$ instead of $2$, one would find $-(6/5)\f\sim 4$, 
but in that case the turn of the fields occurs too near the end of inflation so
that the isocurvature mode will not have disappeared
completely by the end of inflation. Looking at the contributions of $g_{iso}$ 
and $g_{int}$ separately, we see that even in that case $g_{int}$ has already 
gone to a constant while $g_{iso}$ is still decreasing towards zero, 
so that we feel reasonably confident that the estimate is good even
for that model, but we cannot be absolutely certain without a better
treatment of the end of inflation, which is beyond the scope of the present
paper.

\begin{figure}
\begin{center}
\begin{tabular}{cc}
\includegraphics[scale=1]{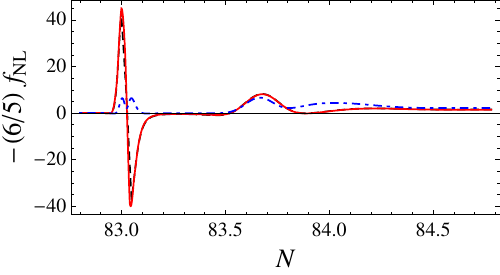}
&\includegraphics[scale=0.9]{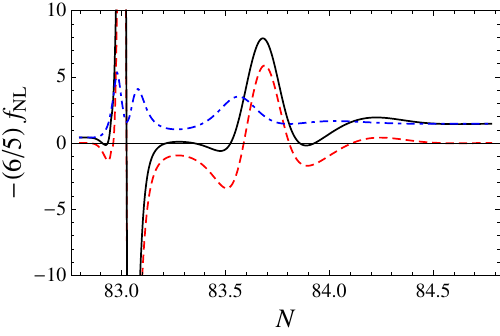}
\end{tabular}
\caption{In the first plot we show the non-Gaussianity parameter $\f$
  as calculated in our formalism exactly (red line) and within the
  analytical slow-roll approximation (blue dot-dashed line) as well as
  numerically in the context of the $\delta N$ formalism (black dashed
  line) for the model (\ref{newpot}). In the second plot we show again
  the total $\f$, now as the black solid line, and split it up into
  the isocurvature contribution proportional to $g_{iso}$ (red dashed
  line) and the integral contribution proportional to $g_{int}$ (blue
  dot-dashed line).  We use the same model as in figure~\ref{fig6}.}
\label{fig8}
\end{center}
\end{figure}

In figure~\ref{fig8} we plot $\f$ for the model (\ref{newpot}) with
the parameter values described above.  Again a notable feature comes
up during the turn of the fields. It comes from the isocurvature term
of (\ref{nuv}) that gets very big during the turn of the field
trajectory, but as soon as the fields relax it vanishes again. We do
not plot $g_{sr}$ separately since it turns out to be negligible. Note
how the final value of $\f$ depends only on the integrated effect, as
the isocurvature contribution has vanished. The final slow-roll
analytical value is calculated to be $-(6/5)f_{\mathrm{NL},sr}=2.15$
while the values obtained numerically by our formalism and the $\delta
N$ formalism are $-(6/5)f_{\mathrm{NL}}=1.43$ and
$-(6/5)f_{\mathrm{NL},\delta N}=1.48$, respectively.  We see excellent
agreement between the exact numerical result of our formalism and the
$\gd N$ one. The slow-roll analytical result does very badly during
the turn of the field trajectory, when slow roll is badly broken, but
gives a reasonable estimate (within $50\%$) of the final value.

\begin{figure}
\begin{center}
\includegraphics[scale=1]{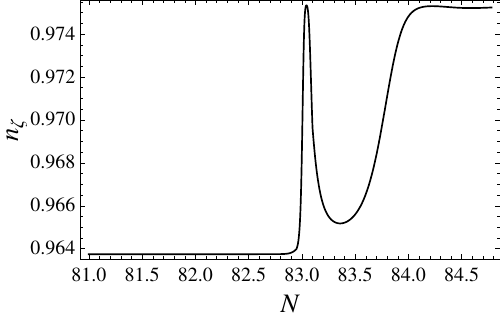}
\caption{The exact numerical spectral index for the same model as in 
figure~\ref{fig6}.}
\label{fig10}
\end{center}
\end{figure}

Finally in figure~\ref{fig10} we plot the spectral index for this
model. Its value is in the range of the $68\%$ confidence levels
after 7 years of WMAP observations \cite{Komatsu:2010fb}, but lies near
the upper limit.

\section{Conclusions}
\label{secConclusion}

The study of the non-Gaussianity produced by inflation models has become a
hot topic of research, since the recent observations of WMAP and in particular
the imminent ones of Planck will allow us to constrain and discriminate 
inflation models based on their non-Gaussian predictions. In this paper we
investigated the super-horizon bispectral non-Gaussianity produced by 
two-field inflation models. To this end we further worked out the 
long-wavelength formalism developed by Rigopoulos, Shellard, and Van Tent
(RSvT) \cite{Rigopoulos:2004ba,Rigopoulos:2005xx,Rigopoulos:2005ae,
Rigopoulos:2005us}. 

We derived an exact result for the bispectrum parameter $\f$ produced
on super-horizon scales for any two-field inflation model with canonical 
kinetic terms, equation~(\ref{fNLresult}). The result is expressed in terms 
of the linear perturbation solutions and slow-roll parameters. However, no
slow-roll approximation has been assumed on super-horizon scales,
these parameters should be viewed as short-hand notation and can be
large. In particular this means that the result is valid for models
where the field trajectory makes a sharp turn in field space so that slow
roll is temporarily broken. On the other hand, we did assume slow roll
to be valid at horizon crossing in order to remove any dependence on the
window function and to use the analytic solutions for the linear mode 
functions. Observations of the scalar spectral index seem to indicate 
that this is a good approximation. Note that the assumption of canonical
kinetic terms is not a fundamental one: the basic equations of the formalism
of RSvT are given for more general kinetic terms. We just did not want to
complicate the notation and expressions in this paper with the covariant
derivatives and additional curvature terms needed to treat the general case.

The result can be split into the sum of three parts, multiplied by an
overall factor (except for a small slow-roll suppressed term that is
the single-field contribution produced at horizon crossing).  This
overall factor is proportional to the contribution of the isocurvature
mode to the adiabatic mode, which is only non-zero for a truly
multiple-field model where the field trajectory makes a turn in field
space, as parametrized by a non-zero value of the slow-roll parameter
$\hpe$. (Effectively) single-field models do not produce any
non-Gaussianity on super-horizon scales, since the adiabatic
perturbation is conserved in that case. The three parts in the sum
are: 1) a part that only involves slow-roll parameters evaluated at
horizon-crossing and hence is always small; 2) a part proportional to
the pure isocurvature mode; and 3) an integral involving terms
proportional to the pure isocurvature mode. Since the adiabatic mode is not
necessarily constant in the presence of isocurvature modes, we only
consider models where the isocurvature mode has disappeared by the end
of inflation, so that we can directly extrapolate our result at the
end of inflation to recombination and observations of the
CMB. However, this automatically means that the part 2), although
varying wildly during the turn of the field trajectory, cannot give
any persistent non-Gaussianity that can be observed in the CMB. This
means that any large non-Gaussianity on super-horizon scales in models
satisfying this condition will have to come from the integrated effect
in part 3).

The exact equation~(\ref{fNLresult}) is the basis of our numerical studies.
However, to gain further insight we tried to work out the integral 
analytically. For this it turns out that the slow-roll approximation is 
necessary. Even then the integral can only be done explicitly for 
certain specific classes of potentials, among which are product potentials,
$W(\gf,\gs) = U(\gf)V(\gs)$, and generalized sum potentials,
$W(\gf,\gs) = (U(\gf)+V(\gs))^\gn$. We found that, with our assumptions
on the disappearance of the isocurvature mode, no product potential can
give large non-Gaussianity, nor can any simple sum potential with equal powers,
$W(\gf,\gs) = \ga \gf^p + \gb \gs^p$. However, we found conditions
under which the (generalized) sum potential can give large non-Gaussianity
(here defined as $\f$ larger than unity), and we have described an explicit, 
simple model that does. It consists of a heavy field rolling down a quadratic 
potential while a light field sits near the local maximum of a double-well 
potential. When the heavy field reaches zero and starts oscillating,
the light field takes over and rolls down, so that there is a turn of the
field trajectory in field space. We studied this model numerically, using the
exact results, to confirm our analytical predictions.

In deriving equation~(\ref{fNLresult}) we assumed that all three scales cross
the horizon at the same moment. However, this is not a necessary assumption,
and we also generalized the result to an arbitrary momentum configuration.
We find that going to the squeezed limit, where one of the momenta is much
smaller than the other two, even when remaining within the resolution
of the Planck satellite ($k' \sim 1000 k$), the result for $\f$ can be 
reduced by about $10\%$, depending on the model. We stress that we are 
discussing $\f$ here, so
this effect is unrelated to the well-known result that the local bispectrum
peaks on squeezed momentum configurations, which is due to the momentum
behaviour of the power spectrum, which has been divided out in $\f$.
However, exactly because of this latter effect, the squeezed limit is very
relevant for the computation of $\f$.

We have worked out and included the second-order source term at horizon 
crossing in the long-wavelength formalism of RSvT, a contribution that
had been missing so far. This is the only change of the basic formalism
with respect to the paper \cite{Rigopoulos:2005us} by RSvT.
While this additional term is always small for the models we
consider and hence numerically insignificant, from an analytical point
of view it means we could now compare our results directly to the so-called
$\f^{(4)}$ as defined in the $\gd N$ formalism 
\cite{Choi:2007su,Byrnes:2008wi,Vernizzi:2006ve}. 
Some of the potentials we studied had already been worked out using
that formalism and where available we compared our analytical results
and found perfect agreement. We also compared our exact numerical
results with those obtained using a numerical $\gd N$ treatment for
models where slow roll is broken and the analytic results cannot be
trusted, and again we found excellent agreement.

We showed that the long-wavelength formalism of RSvT represents a viable
alternative to the $\gd N$ formalism to compute the super-horizon 
non-Gaussianity produced during inflation, allowing us to obtain and
verify results in a different way. Moreover, the long-wavelength formalism
has a number of advantages that can make it preferable in certain situations.
Very importantly, our formalism allows for a simple physical
interpretation of the different parts in terms of adiabatic and
isocurvature modes, providing insight into the behaviour of the different 
transient and persistent contributions to $\f$.
While we did not pursue this in the present paper,
the formalism also provides the solution for the second-order isocurvature
perturbation and hence the isocurvature bispectrum could be computed as
easily as the adiabatic one.

From our studies it has become clear that the condition on the
disappearance of the isocurvature mode by the end of inflation is a
very strong constraint.  It significantly reduces the possibilities
for a large, observable value of $\f$ produced during inflation. 
Note, however, that we chose to impose this condition only to be able to
neglect the further evolution of the adiabatic mode after inflation; it
is in no way a necessary condition for our formalism during inflation.
In future work we would like to relax this condition, which means that
the adiabatic mode would no longer necessarily be constant after
inflation, and hence will require a much better description and
understanding of the evolution of the perturbations during the
transition at the end of inflation and the subsequent period of
(p)reheating. In conclusion, while a lot of progress has been made
over the past few years regarding the non-Gaussianity produced in
multiple-field inflation, more work still remains to be done.

\acknowledgments
BvT would like to thank Gerasimos Rigopoulos and Paul Shellard for many useful
discussions, especially in the initial stages of this work. The authors would
also like to thank Filippo Vernizzi for useful discussions, and S\'ebastien
Renaux-Petel and Thorsten Battefeld for comments on the draft.

\appendix
\label{app}

\section{Basis improvements}
\label{appBasis}

In \cite{GrootNibbelink:2000vx,GrootNibbelink:2001qt} an orthonormal basis in
field space was introduced, with
substantial advantages for calculating and interpreting quantities in
multiple-field inflation (see also \cite{Gordon:2000hv} for a special two-field
case
of this basis). The basis was defined as follows (note that e.g.\ $\vc{\gP}$ is
the vector containing components $\gP^A$). The first basis vector 
$\vc{e}_1$ is the unit vector in the direction of the field velocity. Next, the
direction of the basis vector $\vc{e}_2$ is given by the direction of that part
of the field acceleration that is perpendicular to $\vc{e}_1$. This
orthogonalization process is then continued with higher-order time derivatives,
until a complete basis is found. Defining the generalized $\Bget$ 
parameter as
	\bea
	\Bget^{(n)} \equiv \frac{(\frac{1}{N} \der_t)^{n-1} \BgP}{H^{n-1} \gP}
	\eea 
(with $\Bget \equiv \Bget^{(2)}$ and $\Bgx \equiv 
\Bget^{(3)}$), we can then define the basis vectors via an iterative 
procedure as
	\beq
	\vc{e}_n \equiv \frac{\Bget^{(n)} - \sum_{i=1}^{n-1} 
	\get^{(n)}_i \vc{e}_i}{\get^{(n)}_n}
	\eeq
for $n \geq 2$, with $\vc{e}_1 \equiv \BgP/\gP$ and $\get^{(n)}_i \equiv
\vc{e}_i \cdot \Bget^{(n)}$. Basically there is an arbitrariness in the
choice of sign of the basis vectors, which in the original definition was fixed
by choosing $\get^{(n)}_n$ to be non-negative: 
	\beq
	\get^{(n)}_n \equiv \left | \Bget^{(n)} - \sum_{i=1}^{n-1} 
	\get^{(n)}_i \vc{e}_i \right |.
	\qquad\qquad\mbox{(old definition)}
	\eeq
While being a perfectly valid choice analytically, this choice does mean that
certain basis vector components and slow-roll parameters make sudden sign flips
when one or more fields are oscillating, and that is hard to deal with
numerically. Hence we now propose a different choice for $\get^{(n)}_n$, which
is identical except for the overall sign, and which eliminates the sudden sign
flips:
	\beq
	\get^{(n)}_n \equiv - \gve_{A_1 \cdots A_n} e_1^{A_1} \cdots
	e_{n-1}^{A_{n-1}} \get^{(n) \, A_n},
	\qquad\qquad\mbox{(new definition)}
	\eeq
where $\gve$ is the fully antisymmetric symbol. From the fact that $\Bget
= \sum_{i=1}^n \get^{(n)}_i \vc{e}_i$ it immediately follows that
	\beq
	\gve_{A_1 \cdots A_n} e_1^{A_1} \cdots e_n^{A_n} = -1,
	\eeq
so that this choice means that the basis has a definite handedness. Note that in
the case where the fields do not oscillate, the two definitions have the same
overall sign (hence the choice of the minus sign). To have the expressions for
the time derivative of the basis vectors and the $\get^{(n)}_n$ unchanged, we
see that we also need the relation
	\beq
	\gve_{A_1 \cdots A_n} e_1^{A_1} \cdots e_{n-1}^{A_{n-1}} e_{n+1}^{A_n} 
	= 0
	\eeq
to be satisfied. Then all results and expressions developed with this basis are
unchanged. 

An interesting consequence of these relations, including the orthogonality 
relation
	\beq
	\vc{e}_m \cdot \vc{e}_n = \gd_{mn},
	\eeq
is that for the cases of two and of three fields we have sufficient conditions
to write all basis vectors in terms of $\vc{e}_1$, without knowing anything 
about the dynamics. For two fields we have
	\beq
	\label{e2e1}
	\vc{e}_2 = (e_1^2, -e_1^1),
	\eeq
with $(e_1^1)^2+(e_1^2)^2=1$, and for three fields
	\bea
	\vc{e}_2 & = & (e_1^2+e_1^3, -e_1^1+e_1^3, -e_1^1-e_1^2),\non\\
	\vc{e}_3 & = & \lh {\textstyle\frac{1}{2}} - (e_1^1)^2 - e_1^2 e_1^3,
	-{\textstyle\frac{1}{2}} + (e_1^2)^2 - e_1^1 e_1^3,
	{\textstyle\frac{1}{2}} - (e_1^3)^2 - e_1^1 e_1^2 \rh,
	\eea
with $(e_1^1)^2+(e_1^2)^2+(e_1^3)^2=1$ and $e_1^1 e_1^2-e_1^1 e_1^3+e_1^2 e_1^3
= -\frac{1}{2}$.

\section{Computation of the second-order source term}
\label{appSecondSource}

To compute the second-order source term $b_{ia}^{(2)}$ of
(\ref{secondsource}) we choose a gauge characterized by the constraint
$e_{1A}\varphi^A=0$, so that $\zeta_1=\alpha$ and
$\zeta_2=-(\kappa/\sqrt{2\e})e_{2A}\varphi^A$, where we have split
up $\phi^A(t,\vc{x})=\phi^A(t)+\varphi^A(t,\vc{x})$ and $\ln a(t,\vc{x})
= \ln a(t) + \alpha(t,\vc{x})$. Note that we use here subscripts $1$ and $2$
to indicate the adiabatic and isocurvature components of $\zeta$ (instead
of superscripts).
On long wavelengths this gauge reduces to the uniform energy density gauge. 
We will drop the tilde (that we introduced in section~\ref{secCorrelators} 
to denote quantities in the uniform energy density gauge) and omit the 
superscript $^{(1)}$ on first-order quantities in this appendix in order 
to lighten the notation. 

We can use the momentum constraint (the equivalent of (\ref{fieldeq}) in 
the proper gauge) to solve for the lapse function $N$ to first
order by setting $N=1+N_1$, and we find $N_1=\dot{\zeta}/H$. Following
Maldacena \cite{Maldacena:2002vr} we first write the quadratic action 
in the ADM formalism.
For that we only need to compute $N_1$ to first order. We find that to 
leading order in slow roll the action takes the form
\bea
S_2 = \int \d t \d^3 x L_2 = \int \d t \d^3 x \,
a^3\e\Bigg\{\dot{\zeta}_1^2+\dot{\zeta}_2^2
-4\hpe H\dot{\zeta}_1\zeta_2
+2\chi H\dot{\zeta}_2\zeta_2\Bigg\}.
\eea
To obtain this expression we used the gauge constraint to simplify the
integrand as well as (\ref{srvar}) and (\ref{SRder}).  

For the third-order action 
we find for the leading-order slow-roll terms \cite{Tzavara:2011hn}
\bea
 S_{3}=\int \d t \d^3 x &&\Bigg\{a^3\e^2(\mathrm{interaction\  terms})
-\frac{\delta L_2}{\delta\zeta_2}
\lh-\frac{Q_2}{2}+(\e+\hpa)\zeta_1\zeta_2+\frac{\dot{\zeta}_2\zeta_1}{H}
+\frac{\hpe}{2}\zeta_1^2\rh\nn\\
&&-\frac{\delta L_2}{\delta\zeta_1}\lh-\frac{\zeta_1^{(2)}}{2}
-\frac{Q_1}{2}+\frac{\e+\hpa}{2}\zeta_1^2-\hpe\zeta_1\zeta_2
+\frac{\dot{\zeta}_1\zeta_1}{H}\rh\Bigg\},
\eea
where the auxiliary quantities $Q_i$ are defined as
\be
Q_i=-\frac{H}{\dot{\phi}}e_{iA}\varphi_{(2)}^A,
\ee
and the last two terms of the action are proportional to the second-order
equations of motion. In order for them to vanish we perform a
redefinition of $\zeta_1$ and $\zeta_2$:
\bea\label{gaugeredef}
&&\zeta_1=\zeta_{1c}-\frac{\zeta_1^{(2)}}{2}-\frac{Q_1}{2}
+\frac{\e+\hpa}{2}\zeta_1^2-\hpe\zeta_1\zeta_2
+\frac{\dot{\zeta}_1\zeta_1}{H},\nn\\
&&\zeta_2=\zeta_{2c}-\frac{Q_2}{2}+(\e+\hpa)\zeta_1\zeta_2
+\frac{\dot{\zeta}_2\zeta_1}{H}+\frac{\hpe}{2}\zeta_1^2.
\eea 
The terms $Q_i$ can be rewritten using the definitions of the gradients of 
the curvature perturbations at second order.
While working in the uniform energy density gauge we can use the equivalent
of (\ref{fieldeq}) to find the following constraints to first and second order
\cite{Tzavara:2011hn}:
\bea
&&H\dot{\phi}_A\partial_i\varphi^A_{(1)}=0,\nn\\
&&\frac{1}{2}Q_1=
\frac{\e+\hpa}{2}\zeta_2^2+\frac{\dot{\zeta}_2\zeta_2}{H}
-\partial^{-2}\partial^i\lh\frac{\zeta_2}{H}\partial_i\dot{\zeta}_2\rh,
\label{con}
\eea
while expanding $\zeta_{2i}$ up to second order in the same gauge gives
\bea
\frac{1}{2}\zeta_{2i}^{(2)}&=&\frac{1}{2}\partial_i\zeta_2^{(2)}=
\partial_i\Bigg[\frac{1}{2}Q_2+\frac{1}{H}\zeta_2\dot{\zeta}_1
-\frac{\hpe}{2}\zeta_2^2
-\partial^{-2}\partial^j\lh\frac{\zeta_2}{H}\partial_j\dot{\zeta}_1\rh\Bigg]
\label{z2u}.
\eea  
Then the redefinitions of the gradients become
\bea
\zeta_1+\frac{\zeta_1^{(2)}}{2}=\zeta_{1c}+\frac{\dot{\zeta}_1\zeta_1}{H}
-\frac{\dot{\zeta}_2\zeta_2}{H}
+\frac{\e+\hpa}{2}\lh\zeta_1^2-\zeta_2^2\rh-\hpe\zeta_1\zeta_2
+\partial^{-2}\partial^i\lh\frac{\zeta_2}{H}\partial_i\dot{\zeta}_2\rh,
\eea
\bea
\zeta_2+\frac{\zeta_2^{(2)}}{2}=\zeta_{2c}+\frac{\zeta_2\dot{\zeta}_1}{H}
+\frac{\zeta_1\dot{\zeta}_2}{H}+\frac{\hpe}{2}\lh\zeta_1^2-\zeta_2^2\rh
+(\e+\hpa)\zeta_1\zeta_2
-\partial^{-2}\partial^i\lh\frac{\zeta_2}{H}\partial_i\dot{\zeta}_1\rh.
\eea
Inspecting (\ref{redefinedzetagauge}) and (\ref{z2u}) we see that in the 
uniform energy density gauge $\zeta_{mi}=\partial_i\zeta_m$ both for the 
adiabatic and the isocurvature component. Since we perform our main 
calculation in the flat gauge and we use the variable 
$\zeta_{mi}$ rather than $\zeta_m$, we want to transform the 
above redefinitions to this gauge by the simple gauge transformation
(\ref{mw}):
\be
\zeta_{mi}^{(2)}=\tilde{\zeta}_{mi}^{(2)}-\frac{\zeta_1}{H}\dot{\zeta}_{mi},
\ee
where we used again the tilde to denote the uniform energy density gauge.
We find
\be
\zeta_{1i}+\frac{\zeta_{1i}^{(2)}}{2}=\partial_i\Big[\zeta_{1c}
+\frac{\dot{\zeta}_1\zeta_1}{H}
-\frac{\dot{\zeta}_2\zeta_2}{H}
+\frac{\e+\hpa}{2}\lh\zeta_1^2-\zeta_2^2\rh-\hpe\zeta_1\zeta_2
+\partial^{-2}\partial^j\lh\frac{\zeta_2}{H}\partial_j\dot{\zeta}_2\rh\Bigg]
-\frac{\zeta_1}{H}\partial_i\dot{\zeta}_1
\label{z1fg}
\ee
\be
\zeta_{2i}+\frac{\zeta_{2i}^{(2)}}{2}=\partial_i\Big[\zeta_{2c}
+\frac{\zeta_2\dot{\zeta}_1}{H}
+\frac{\zeta_1\dot{\zeta}_2}{H}+\frac{\hpe}{2}\lh\zeta_1^2-\zeta_2^2\rh
+(\e+\hpa)\zeta_1\zeta_2
-\partial^{-2}\partial^j\lh\frac{\zeta_2}{H}\partial_j\dot{\zeta}_1\rh
\Bigg]-\frac{\zeta_1}{H}\partial_i\dot{\zeta}_2\label{z2f}
\ee
Note that after horizon exit $\dot{\zeta}_1=2H\hpe\zeta_2$ and 
$\dot{\zeta}_2=-H\chi\zeta_2$ (the latter is valid under the slow-roll 
approximation only) in the 
gauge used in this appendix, so that the expressions simplify.
For a field redefinition of the form $\zeta=\zeta_c+\lambda\zeta_c^2$ 
(note that in the equations above we have not added the subscript $c$
explicitly in the quadratic terms, since to second order it makes no 
difference) the three-point correlation function can be written as
\be
\langle\zeta\zeta\zeta\rangle=\langle\zeta_c\zeta_c\zeta_c\rangle
+2\lambda[\langle\zeta_c\zeta_c\rangle
\langle\zeta_c\zeta_c\rangle+\mathrm{cyclic}].
\ee
Hence the elements of $L_{1ab}$ and $N_{1ab}$ are just the 
coefficients of the various products of $\zeta_1$ and $\zeta_2$ in the 
redefinition of $\zeta_1$ multiplied by $2$, and similarly for 
$L_{2ab}$ and $N_{2ab}$. Note that the local terms $L_{abc}$ correspond
to the terms between the square brackets, and the non-local terms $N_{abc}$ 
to the terms outside. This leads to the explicit slow-roll expressions
given at the end of section~\ref{secGreen}.

\section{Gauge issues}
\label{appGauge}

\subsection{Gauge invariant quantities}

The use of spatial gradients was first advocated in \cite{Ellis:1989jt} in the
context of the covariant formalism. Later on the authors of
\cite{Rigopoulos:2005xx} constructed invariant quantities under long-wavelength
changes of time-slicing, by considering the
following combination of the spatial gradients of two spacetime scalars $A$ 
and $B$:
\be
\mathcal{C}_i\equiv \partial_iA-\frac{\partial_tA}{\partial_tB}\partial_iB,
\ee
and formed among others the quantity
\be
\zeta_i=\partial_i\alpha-\frac{\partial_t\alpha}{\partial_t\rho}\partial_i\rho,
\ee
which when linearized is just the gradient of the comoving curvature
perturbation $\zeta$. Here $\ga$ is the logarithm of the scale factor and 
$\rho$ the energy density.
In \cite{Langlois:2005qp} it was argued that this particular
combination could give rise to a second-order gauge-invariant quantity
$\gz^{(2)}$ as follows:
\be
\partial_i\zeta^{(2)}=\zeta_i^{(2)}-\frac{\rho^{(1)}}{\dot{\rho}}
\dot{\zeta}^{(1)}_i,
\label{mw}
\ee
where $\zeta^{(2)}$ can be related to yet another gauge-invariant second-order
quantity defined by Malik and Wands in \cite{Malik:2003mv},
\be
\zeta^{(2)}\simeq \zeta^{(2)}_{MW}-\zeta^{(1)2}_{MW}.
\ee

We can easily show that (\ref{mw}) is equivalent to taking the gauge
transformation of $\zeta^{(2)1}_i$ from homogeneous expansion
time slices to uniform energy density time slices. Denoting as
tilded the quantities in the new time slices, where $T=t+\Delta t$ (no
space transformation is required), we can write
\bea
\tilde{\zeta}_i^{(1)1}&=&\zeta_i^{(1)1},\nn\\
\tilde{\zeta}_i^{(2)1}&=&\zeta_i^{(2)1}-\Delta t\ \dot{\zeta}_i^{(1)1}.
\eea
The quantity $\Delta t$ is the time difference between the two time slices and
can be evaluated by either comparing $\alpha$ (see also (\ref{gaugetrans})) 
or $\rho$ in the two gauges:
\be
\rho(t,x)=\tilde{\rho}(t+\Delta t)=\tilde{\rho}(t)+\Delta
t\ \dot{\tilde{\rho}}(t),
\ee
which expanded to first order becomes
\be
\rho^{(0)}(t)+\rho^{(1)}(t,x)=\tilde{\rho}(t)+\Delta
t\ \dot{\tilde{\rho}}(t).
\ee
Hence we find that $\rho^{(0)}=\tilde{\rho}$ and
\be
\Delta t=\frac{\rho^{(1)}(t,x)}{\dot{\tilde{\rho}}(t)}.
\ee
We conclude that our definition of the second order $\tilde{\zeta}^{(2)1}_i$ 
in the uniform energy density gauge agrees with the gauge-invariant 
quantities defined by other authors. 

\subsection{Gradients and locality}

As a consistency check we want to verify that $\tilde{\zeta}^{(2)1}_i$ is
indeed a total gradient, as it should be according to the first part of this
appendix and (\ref{redefinedzetagauge}). Taking expression
(\ref{u2hor}) (corrected by the gauge transformation),
\bea
 \tilde{\zeta}^{(2)1}_i\!\!&=&\!\!-(\partial_iv_{e*})v_{f*}
\Bigg\{\!\!-2\hpe G_{1f}(t,t_*)G_{2e}(t,t_*)
+\int_{t_*}^t\!\!\mathrm{d} t' G_{1a}(t,t')
\bar{A}_{abc}G_{be}(t',t_*)G_{cf}(t',t_*)\Bigg\}\nn\\
 &&\!\!+G_{1a}(t,t_*)L_{aef*}\partial_i(v_{e*}v_{f*})+G_{1a}(t,t_*)N_{aef*}(\partial_iv_{e*})v_{f*},
\eea
and rewriting it using (\ref{timederA}) we find
\bea
 \tilde{\zeta}^{(2)1}_i\!\!&=&\!\!-(\partial_iv_{e*})v_{f*}
\Bigg\{\!\!A_{ab}(t_*)G_{1a}(t,t_*)\delta_{be}\delta_{1f}
+\int_{t_*}^t\!\!\mathrm{d} t' G_{1a}(t,t')
\bar{A}_{ab\bar{c}}G_{be}(t',t_*)G_{\bar{c}f}(t',t_*)\nn\\
 &&\qquad\qquad\ \ \  -\int_{t_*}^t\!\!\mathrm{d}t'A_{ab}A_{1c}G_{1a}(t,t')G_{be}(t',t_*)G_{cf}(t',t_*)-G_{1a}(t,t_*)N_{aef*}\Bigg\}\nn\\
 &&\!\!+G_{1a}(t,t_*)L_{aef*}\partial_i(v_{e*}v_{f*}).\label{zu}
\eea
This expression should be symmetrical under the interchange of the indices 
$e$ and $f$. Notice that the last term is automatically symmetrical.

The anti-symmetrical
part of the two integrands turns out to be proportional to 
\be
T_a=G_{23}(t',t_*)G_{32}(t',t_*)-G_{22}(t',t_*)G_{33}(t',t_*).
\label{sym}
\ee
We explicitly check the exact numerical value of this quantity and find
it to be zero. For this it is crucial that we have defined $t_*$ as the time
a few (about 3) e-folds after horizon crossing.
The reason is that the long-wavelength 
approximation we use in all our derivations is only valid once the rapidly
decaying mode can be neglected, which takes a few e-folds. If the above
quantity were to be evaluated before that time, it would not yet be zero.
Note that in the slow-roll case $T_a$ is identically zero according to 
(\ref{GsolSR}), since within the slow-roll approximation the decaying mode 
is neglected by construction.
The above means that $\tilde{\zeta}_i^{1}$ is well defined only after it is 
well outside the horizon, where we can neglect the decaying mode. 
The case where the decaying mode can remain important is treated in the 
one-field case in \cite{Takamizu:2010xy}.

The remaining non-integral terms between the braces can be explicitly checked 
to cancel when taking the slow-roll limit at horizon crossing: the first term 
of the first line of (\ref{zu}) gives
\be
-(\partial_iv_{e*})v_{f*}
A_{ab}(t_*)G_{1a}(t,t_*)\delta_{be}\delta_{1f}=\lh2\hpe
-\chi G_{12}(t,t_*)\rh\lh v_{1*}\partial_iv_{2*}-v_{2*}\partial_iv_{1*}\rh,
\ee
while the terms arising from the non-local contribution $N_{aef*}$ gives 
exactly the same but with opposite sign. Hence we see that within the
conditions of the long-wavelength approximation $\tilde{\zeta}^{(2)1}_i$
is indeed a total gradient, as it should be.

\section{Detailed calculations}

\subsection{Relation between space and time derivatives}
\label{appTimeder}

We begin by proving that
	\beq\label{timederA}
	\bA_{ab1} = - \frac{1}{NH} \der_t A_{ab}
	\eeq
for all gauges with $\der_i \ln a = 0$. Actually the statement is more general:
the prefactor of $\gz_i^1$ in the expression for $\der_i f$, where $f$ is any
function of $H,\Bgf,\BgP$, is equal to $-(\der_t f)/(NH)$ for gauges satisfying
$\der_i \ln a = 0$.

We start by computing the time derivative of $f$:
	\bea
	\frac{\der_t f(H,\Bgf,\BgP)}{NH} & = & \frac{1}{NH} 
	\lh \der_H f \, \der_t H + \Bnabla_\Bgf f \cdot
	\der_t \Bgf + \Bnabla_\BgP f \cdot \der_t \BgP \rh \non\\
	& = & -H \ge \, \der_H f 
	+ \frac{\sqrt{2\ge}}{\gk} \, \vc{e}_1 \cdot \Bnabla_\Bgf f
	+ \frac{\sqrt{2\ge}}{\gk} H \Bget \cdot \Bnabla_\BgP f,
	\eea
where we used the definitions of the slow-roll parameters $\ge$ and
$\Bget$, and of the basis vector $\vc{e}_1$.
On the other hand, the spatial derivative of $f$, in a $\der_i \ln a = 0$ gauge,
is given by
	\bea
	\der_i f(H,\Bgf,\BgP) & = &  
	\der_H f \, \der_i H + \Bnabla_\Bgf f \cdot
	\der_i \Bgf + \Bnabla_\BgP f \cdot \der_i \BgP \non\\
	& = & H \ge \, \der_H f \, \vc{e}_1 \cdot \Bgz_i
	- \frac{\sqrt{2\ge}}{\gk} \, \Bgz_i \cdot \Bnabla_\Bgf f\\
	&&- \frac{\sqrt{2\ge}}{\gk} H \lh {\textstyle\frac{1}{NH}} \, \Bgth_i
	+(\ge+\get^\parallel)\Bgz_i -\ge (\vc{e}_1 \cdot \Bgz_i)
	\vc{e}_1 \rh \cdot \Bnabla_\BgP f, \non
	\eea
using the constraint relations for $\der_i H$, $\der_i \Bgf$, and $\der_i \BgP$
given in \cite{Rigopoulos:2005xx,Rigopoulos:2005us}. (Note that some time
and space derivatives have to be replaced with their covariant (in field space)
version to make contact with the more general expressions given in those papers
that take into account a non-trivial field metric.)
Taking the components of $\Bgz$ and $\Bgth$ in the basis defined in 
\ref{appBasis}, not forgetting the
relation $\vc{e}_m \cdot \Bgth_i = \gth_i^m + NH Z_{mn} \gz_i^n$, with
$Z_{21}=\get^\perp$ \cite{Rigopoulos:2005xx,Rigopoulos:2005us}, we prove the
stated relation, of which (\ref{timederA}) is a special case.

\subsection{Derivation of equation (\ref{fNLresult})}
\label{derivation}

In this appendix we work out the last term of (\ref{fNLint}), which has to
be added to the result for the second term ($I$) given in (\ref{Ires}), to
derive the final expression (\ref{fNLresult}) for $\f$. We call the sum of
these two terms $J$: 
\beq
\frac{J}{\gamma_*^2} \equiv \frac{I}{\gamma_*^2} 
+ \int_{t_*}^t \d t' G_{1a} \bA_{ab\bc} \bv_{bm} \bv_{\bc n},
\eeq
which is
        \bea
	\frac{J}{\gamma_*^2} = & \gd_{m2} \gd_{n1} 
	\Big( -2 \get^\perp_* + & \gc_* G_{12}(t,t_*)+ A_{32*} G_{13}(t,t_*)
	- \gc_* A_{33*} G_{13}(t,t_*) \Big) \nn\\
	&- \gd_{m2} \gd_{n2} \, 2 \get^\perp_* \gc_*
	& G_{13}(t,t_*) \\
 & + \gd_{m2} \gd_{n2} \int_{t_*}^t \d t' \Big [ & 
\lh \bA_{122} - 4(\get^\perp)^2 + \bA_{322} G_{13} \rh (\bv_{22})^2
+ \lh \bA_{333} + 2\get^\perp \rh G_{13} (\bv_{32})^2 \nn\\
	&&  + \lh \bA_{123} - 4\get^\perp G_{12} 
	+ (\bA_{323} + \bA_{332}
	+ 2\get^\perp A_{33} + 2 \dot{\get}^\perp) G_{13} \rh 
	\bv_{22} \bv_{32} 
	\Big] . \non
	\eea
We remind the reader that the bar on top of an index ($\bc$) means that it 
does not take the value 1 and that a subscript $*$ means that a quantity is
evaluated at $t_*$.
The explicit form of the matrix $\mx{\bA}$ is given in \cite{Rigopoulos:2005us}:
\bea
\bA_{121} & = & 2\ge\get^\perp
-4\get^\parallel\get^\perp + 2\gx^\perp, \non\\
\bA_{122} & = & -6\gc - 2\ge\get^\parallel - 2(\get^\parallel)^2 
- 2(\get^\perp)^2, \non\\
\bA_{123} & = &	-6 - 2\get^\parallel, \non\\
\bA_{321}&=&-12\ge\get^\parallel - 12(\get^\perp)^2 
- 6\ge\gc - 8\ge^3 - 20\ge^2 \get^\parallel 
- 4\ge(\get^\parallel)^2 - 12\ge(\get^\perp)^2\nn\\
&&+ 16\get^\parallel(\get^\perp)^2 - 6\ge\gx^\parallel
- 12\get^\perp\gx^\perp + 3 (\tilde{W}_{111}-\tilde{W}_{221}),\nn\\
\bA_{322}&=&-24\ge\get^\perp - 12\get^\parallel\get^\perp 
+ 24\get^\perp\gc - 12\ge^2\get^\perp + 8(\get^\parallel)^2\get^\perp 
+ 8(\get^\perp)^3\nn\\
&& - 8\ge\gx^\perp - 4\get^\parallel\gx^\perp
+ 3 (\tilde{W}_{211}-\tilde{W}_{222}),\nn\\
\bA_{323}&=& 12\get^\perp - 4\ge\get^\perp + 8\get^\parallel\get^\perp
- 4\gx^\perp, \non\\
\bA_{331} & = & -2\ge^2 - 4\ge\get^\parallel +
2(\get^\parallel)^2 - 2(\get^\perp)^2 - 2\gx^\parallel, \non\\
\bA_{332} & = &	-4\ge\get^\perp - 2\gx^\perp, \non\\ 
\bA_{333} & = & -2\get^\perp,
\label{Abar}
\eea
while the rest of the matrix elements are zero.
Using these expressions we have
	\bea
	\bA_{333} + 2 \get^\perp = 0,
	\non\\
	\bA_{323} + \bA_{332} + 2 \get^\perp A_{33} + 2 \dot{\get}^\perp
	= 18 \get^\perp - 4 \dot{\get}^\perp,
	\non\\
	\bA_{123} = -2 A_{33} + 2 \ge + 2 \get^\parallel,
	\non\\
	\bA_{122} - 4 (\get^\perp)^2 
	= -2 A_{32} + 2 \dot{\ge} + 2 \dot{\get}^\parallel,
	\eea
so that we can write
	\bea
	\frac{J}{\gamma_*^2} = & \gd_{m2} \gd_{n1} 
	\Big( -2 \get^\perp_* + & \gc_* G_{12}(t,t_*)+ A_{32*} G_{13}(t,t_*)
	- \gc_* A_{33*} G_{13}(t,t_*) \Big) \nn\\
	&- \gd_{m2} \gd_{n2} \, 2 \get^\perp_* \gc_*
	& G_{13}(t,t_*) \non\\
 & + \gd_{m2} \gd_{n2} \int_{t_*}^t \d t' \Big [ & 
	2 (\bv_{22})^2  \frac{\d}{\d t'}(\ge+\get^\parallel)
	+ 2 \bv_{22} \frac{\d}{\d t'} \bv_{32}
	- 4 \lh \get^\perp G_{12} + \dot{\get}^\perp G_{13} \rh
	\frac{1}{2} \frac{\d}{\d t'} (\bv_{22})^2 \non\\
 && + 2 (\ge+\get^\parallel) \bv_{22} \bv_{32}
	+ \bA_{322} G_{13} (\bv_{22})^2
	+ 18\get^\perp G_{13} \bv_{22} \bv_{32} 
	\Big ].
	\eea
Doing integrations by parts on the three terms in the third line we obtain
	\bea
	\frac{J}{\gamma_*^2} = & \gd_{m2} \gd_{n1} 
	\Big( -2 & \get^\perp_* + \gc_* G_{12}(t,t_*)+ A_{32*} G_{13}(t,t_*)
	- \gc_* A_{33*} G_{13}(t,t_*) \Big) \nn\\
	& + 2 \gd_{m2} \gd_{n2} \Biggl (
	& - \get^\perp_* \gc_* G_{13}(t,t_*)
	- (\ge_* + \get^\parallel_*) + \gc_*
	+ \get^\perp_* G_{12}(t,t_*)\nn\\
 	&&
	+ \dot{\get}^\perp_* G_{13}(t,t_*)
	+ (\ge + \get^\parallel) (\bv_{22})^2
	+ \bv_{22} \bv_{32} \Biggr ) \nn\\
	& + 2 \gd_{m2} \gd_{n2} & \int_{t_*}^t \d t' \Big [ 
	- 2 (\get^\perp)^2 (\bv_{22})^2
	- (\ge+\get^\parallel) \bv_{22} \bv_{32} 
	- (\bv_{32})^2
	+ 9 \get^\perp G_{13} \bv_{22} \bv_{32}\nn\\
	&& \qquad\quad
	+ \frac{1}{2} \lh \bA_{322} + 2 \ddot{\get}^\perp 
	+ 2 \dot{\get}^\perp A_{33} + 2 \get^\perp A_{32} 
	\rh G_{13} (\bv_{22})^2 \Big ] .
	\eea

The following relation (derived by taking two time derivatives of the 
field equation) can be used to remove higher-order slow-roll parameters:
	\beq\label{Wm11rel}
	\tilde{W}_{m11}
	= -\frac{\get^{(4)}_m}{3} - \lh 1-\frac{\get^\parallel}{3} \rh \gx_m
	+ (2\ge+\get^\parallel) \get_m + \ge \get^\parallel \gd_{m1}
	- \get^\perp \tilde{W}_{m2}.
	\eeq
Explicitly, for $m=1$ and $m=2$ in the case of two fields, this becomes
	\bea\label{W111W211rel}
	\tilde{W}_{111}
	& = & -\frac{1}{3} \get^{(4)\,\parallel} 
	- \lh 1-\frac{1}{3}\get^\parallel \rh \gx^\parallel
	+ 3\ge\get^\parallel + (\get^\parallel)^2 + (\get^\perp)^2
	+ \frac{1}{3} \get^\perp \gx^\perp,
	\non\\
	\tilde{W}_{211}
	& = & -\frac{1}{3} \get^{(4)\,\perp} 
	- \lh 1-\frac{1}{3}\get^\parallel \rh \gx^\perp
	+ 3\ge\get^\perp + 2 \get^\parallel \get^\perp
	- \get^\perp \gc.
	\eea
Using the second of these relations, as well as the explicit expression for 
$\bA_{322}$, we find that
\bea
	&&\bA_{322} + 2 \ddot{\get}^\perp + 2 \dot{\get}^\perp A_{33}
	+ 2 \get^\perp A_{32}\\
&&\qquad\qquad	
        \!\!=24 \get^\perp \gc - 12 \get^\parallel \get^\perp
	+ 12 (\get^\parallel)^2 \get^\perp
        + 12 (\get^\perp)^3
	- 4 \get^\perp \gx^\parallel - 4 \get^\parallel \gx^\perp
	- 3(\tilde{W}_{211} + \tilde{W}_{222}). \non
	\eea
We now drop boundary terms that are second order in the slow-roll parameters 
{\em at horizon crossing}, since it would be inconsistent to include them given
that the linear solutions used at horizon crossing are only given up to first 
order. Then the result is
	\bea
	\frac{J}{\gamma_*^2} = & \gd_{m2} \gd_{n1} 
	 ( -2 \get^\perp_* + \gc_* & \bv_{12} ) 
	+ 2 \gd_{m2} \gd_{n2} \lh
	- \ge_*\! - \get^\parallel_*\! + \gc_*
	+ \get^\perp_* \bv_{12}
	+ (\ge + \get^\parallel) (\bv_{22})^2
	+ \bv_{22} \bv_{32} \rh \nn\\
 & + 2 \gd_{m2} \gd_{n2} \int_{t_*}^t \d t' \Bigg [ 
	& - 2 (\get^\perp)^2 (\bv_{22})^2
	- (\ge+\get^\parallel) \bv_{22} \bv_{32}
	- (\bv_{32})^2 + 9 \get^\perp G_{13} \bv_{22} \bv_{32} \non\\
 &&  
	+ \Bigg( 12 \get^\perp \gc - 6 \get^\parallel \get^\perp
	+ 6 (\get^\parallel)^2 \get^\perp  + 6 (\get^\perp)^3
	- 2 \get^\perp \gx^\parallel - 2 \get^\parallel \gx^\perp \non\\
 && \qquad
	- \frac{3}{2}( \tilde{W}_{211} + \tilde{W}_{222})
	\Bigg) G_{13} (\bv_{22})^2 \Bigg ] .
	\eea
Inserting this into (\ref{fNL}) gives the final result for $\f$ in
(\ref{fNLresult}).

\bibliography{bib}{}

\bibliographystyle{JHEP.bst}

\end{document}